\documentclass[useAMS,usenatbib]{mn2e}

\usepackage{graphicx}
\usepackage{natbib}
\usepackage[figuresleft]{rotating}
\newcommand{\HI}{H{\,\small I}}
\newcommand{\ltsima} {$\; \buildrel < \over \sim \;$}
\newcommand{\gtsima} {$\; \buildrel > \over \sim \;$}
\newcommand{\lta} {\lower.5ex\hbox{\ltsima}}
\newcommand{\gta} {\lower.5ex\hbox{\gtsima}}
\newcommand{\kmsMp}{km s$^{-1}$ Mpc$^{-1}$}
\newcommand{\kms}{km\ s$^{-1}$}
\newcommand{\FRI}{FR{-\small I}}
\newcommand{\FRII}{FR{-\small II}}
\newcommand{\paperI}{Paper{\,\small I}}
\title[H{\tiny \ }{\small I} in nearby radio galaxies]{Large-scale HI in nearby radio galaxies (II): the nature of classical low-power radio sources}
\author[B. H. C. Emonts et al.]{B. H. C. Emonts$^{1}$\thanks{E-mail:bjorn.emonts@csiro.au}\thanks{Bolton Fellow}, R. Morganti$^{2,3}$, C. Struve$^{2,3}$, T. A. Oosterloo$^{2,3}$,
\newauthor G. van Moorsel$^{4}$, C. N. Tadhunter$^{5}$, J. M. van der Hulst$^{3}$, E. Brogt$^{6,7}$, 
\newauthor J. Holt$^{8}$, N. Mirabal$^{9,10}$\\
$^{1}$Australia Telescope National Facility, CSIRO Astronomy and Space Science, PO Box 76, Epping NSW, 1710, Australia\\ 
$^{2}$Netherlands Institute for Radio Astronomy, Postbus 2, 7990 AA Dwingeloo, the Netherlands\\
$^{3}$Kapteyn Astronomical Institute, University of Groningen, P.O. Box 800, 9700 AV Groningen, the Netherlands\\
$^{4}$National Radio Astronomy Observatory, Socorro, NM 87801, USA\\ 
$^{5}$Department of Physics and Astronomy, University of Sheffield, Sheffield S3 7RH, UK\\
$^{6}$The University of Arizona, Steward Observatory, 933 North Cherry Avenue, Tucson, AZ 85721, USA\\
$^{7}$University Centre for Teaching and Learning, University of Canterbury, Private Bag 4800, Christchurch 8140, New Zealand\\
$^{8}$Leiden Observatory, Leiden University, PO Box 9513, 2300 RA Leiden, The Netherlands\\
$^{9}$Department of Astronomy, Columbia University,  Mail Code 5246, 550 West 120th Street, New York, N.Y. 10027, USA\\ 
$^{10}$Dpto. de F\'{i}sica Atomica, Molecular y Nuclear, Universidad Complutense de Madrid, E-28040 Madrid, Spain
}
\begin{document}

\date{}

\pagerange{\pageref{firstpage}--\pageref{lastpage}} \pubyear{2007}

\maketitle

\label{firstpage}

\begin{abstract}
An important aspect of solving the long-standing question as to what triggers various types of Active Galactic Nuclei involves a thorough understanding of the overall properties and formation history of their host galaxies. This is the second in a series of papers that systematically study the large-scale properties of cold neutral hydrogen (\HI) gas in nearby radio galaxies. The main goal is to investigate the importance of gas-rich galaxy mergers and interactions among radio-loud AGN. In this paper we present results of a complete sample of classical low-power radio galaxies. We find that extended Fanaroff $\&$ Riley type-I radio sources are generally not associated with gas-rich galaxy mergers or ongoing violent interactions, but occur in early-type galaxies without large ($\ga 10^8 M_{\odot}$) amounts of extended neutral hydrogen gas. In contrast, enormous discs/rings of \HI\ gas (with sizes up to 190 kpc and masses up to $2 \times 10^{10} M_{\odot}$) are detected around the host galaxies of a significant fraction of the compact radio sources in our sample. This segregation in \HI\ mass with radio source size likely indicates that these compact radio sources are either confined by large amounts of gas in the central region, or that their fuelling is inefficient and different from the fuelling process of classical \FRI\ radio sources. To first order, the overall \HI\ properties of our complete sample (detection rate, mass and morphology) appear similar to those of radio-quiet early-type galaxies. If confirmed by better statistics, this would imply that low-power radio-AGN activity may be a short and recurrent phase that occurs at some point during the lifetime of many early-type galaxies.
\end{abstract}

\begin{keywords}

ISM: kinematics and dynamics -- galaxies: active -- galaxies: evolution -- galaxies: interactions --  galaxies: jets -- radio lines: galaxies

\end{keywords}

\section{Introduction}
\label{sec:intro}
Active Galactic Nuclei (AGN) are believed to be triggered when gas and matter are deposited onto a super-massive black hole in the centre of the host galaxy. For this to happen, the gas needs to lose sufficient angular momentum to be transported deep into the potential well of the galaxy, until it eventually fuels the AGN. Many different mechanisms have been proposed for transporting the gas down to the nuclear region, from galaxy mergers and interactions \citep[e.g.][]{hec86,lin88,wu98,col95,can01,kuo08} to bars and central spiral structures \citep*[e.g.][]{sch89,pri05} to cooling flows and accretion of circum-galactic hot gas \citep[e.g.][]{fab95,bes06,all06}. Undoubtedly, many of these fuelling mechanisms do occur; since various classes of AGN (e.g quasars, Seyferts, radio galaxies, etc.) are found in different environments and are known to have intrinsic differences \citep[other than simply orientation-dependent properties, see e.g.][]{urr95}, it is likely that certain mechanisms are associated with specific types of AGN \citep[see e.g.][for a review]{mar04}.

Nearby radio galaxies form a particularly interesting group of active galaxies for investigating possible AGN fuelling mechanisms. Their radio continuum sources evolve over time and both fuelling characteristics as well as dynamical interaction with their surrounding host galaxies can reflect in easily observable properties of these sources. This allows to estimate the time-scale since the onset of the current episode of AGN activity as well as to match radio source characteristics with host galaxy properties and possible fuelling mechanisms. 

For example, compact radio sources (in particular the Giga-hertz Peaked Spectrum [GPS] and Compact Steep Spectrum [CSS] sources) are often believed to be young radio sources. Interactions between their radio jets (which can be imaged at high resolution with VLBI observations) and the surrounding medium can give an insight into the physical properties of the Inter-Stellar Medium (ISM) in the central region of the galaxy, where the potential AGN fuel reservoir is stored. 

On larger scales, there is a striking dichotomy in radio source morphology between high- and low-power radio sources \citep{fan74}. While powerful, Fanaroff $\&$ Riley type-II (\FRII), radio sources contain relativistic jets that end in bright hot-spots, low-power \FRI\ sources are sub-relativistic and have an edge-darkened morphology. Various studies indicate that this striking difference in radio-source properties may be linked to a difference in host galaxy properties and, related, difference in the feeding mechanism of the AGN. It has been argued from optical studies that a significant fraction of powerful radio galaxies with strong emission-lines show peculiar optical morphologies and emission-line features reminiscent of a gas-rich galaxy merger, but that low-power radio sources with weak emission-lines do not generally share the same optical properties \citep{hec86,bau92}. \citet*{chi99} show from {\sl HST} observations that low-power radio sources lack evidence for an obscuring torus and substantial emission from a classical accretion disc. This suggests that accretion may take place in a low efficiency regime, which can be explained by accretion of gas from the galaxy's hot gaseous halo \citep{fab95}. From X-ray studies, \citet*{har07} suggest that high-excitation AGN in general (comprising a large fraction of powerful radio sources) may form a classical accretion disc from cold gas deposited by a gas-rich galaxy merger, while low-excitation AGN (comprising most low-power radio galaxies) may be fed through a quasi-spherical Bondi accretion of circum-galactic hot gas that condenses directly onto the central black-hole. A similar conclusion was reached by \citet{bal08} from the fact that high-excitation radio galaxies almost always show evidence for recent star formation, while this is generally not the case in their low-excitation counterparts. 

Interestingly, all the above mentioned studies place an important emphasis on the crucial role of the {\sl cold gas}, since this gas -- when deposited after a gas-rich galaxy merger/collision -- is thought to be the potential fuel reservoir for AGN/starburst activity and is also believed to be be an important ingredient in the formation of a classical accretion disc and surrounding torus. Unfortunately, so far a systematic inventory of the cold gas in radio galaxies is crucially lacking.

Studying the neutral hydrogen (\HI) gas in radio galaxies provides a powerful tool to investigate the occurrence of cold gas among various types of radio-loud AGN. {\sl \HI\ observations are particularly suited to reveal the occurrence of gas-rich galaxy mergers and interaction in these systems and hence study their importance in triggering/fuelling the radio source.} This issue was first addressed by \citet{hec83}, who used single-dish \HI\ observations for tracing the cold gas in a pre-selected sample of radio-loud interacting galaxies. Current-day interferometers allow us to map the \HI\ gas in unbiased samples of nearby radio galaxies. Mapping the \HI\ in emission on host galaxy scales can reveal ongoing galaxy interactions that are easily missed by optical imaging of the starlight; see for example the case of the M81 group \citep*[][]{Yun94}. A good example of this is also given by \citet{kuo08} and \citet{tan08}, who show that \HI\ observations of nearby Seyfert galaxies clearly reveal that Seyfert systems are much more strongly associated with ongoing interactions than their non-active counterparts -- a trend that is not seen from optical analysis of their samples. 

In case of a more violent galaxy merger or collision, the simultaneous spatial and kinematical information obtained with \HI\ observations is ideal for tracing and dating these events over relatively long time-scales. The reason is that in a galaxy merger or collision, part of the gas is often expelled in the form of large structures \citep[tidal tails, bridges, shells, etc.;][]{hib96,mih96,bar02}, which often have a too low surface density for massive star formation to occur. If the environment is not too hostile, parts of these gaseous structures remain bound to the host galaxy as relic signs of the galaxies' violent past, even long after optical stellar features directly associated with this encounter may have faded \citep[e.g.][]{hib96}. Several studies show that time-delays of tens to many hundreds of Myr between a merger event and the onset of the current episode of AGN activity do occur among active galaxies \citep{tad05,emo06,lab08}, making \HI\ observations ideal for detecting evidence of galaxy mergers or collisions on these time-scales. 

\begin{table*}
\centering
\caption{Radio galaxy properties}
\vspace{0.4cm}
\label{tab:sourceproperties}
\begin{tabular}{lllcccccclclc}
Source  & & \multicolumn{1}{c}{{\sl z}} & D & Opt.  & $M_{V}$  & S(60$\mu$m) & S(100$\mu$m) & log$P_{\rm 1.4 GHz}$ & (ref.) & LS & (ref.) & Type  \\
B2 name & other name & & (Mpc) & Mor. &          & (mJy)       & (mJy) & (W/Hz) & & (kpc) & & source  \\
\hline
0034+25	&          &0.0318	&134          & E         & -22.1  & $<$153      & $<$378      &	23.4 & (3) & 200  & (5,10) & \FRI\  \\
0055+30 & NGC 315  & 0.0165     & 70          & E         & -23.1  & 363$\pm$17  & 586$\pm$74  & 24.2 & (6) & 1200 & (11,15)   & \FRI\  \\
0104+32 & 3C31     & 0.0169     & 71          & S0        & -22.0  & 444$\pm$21  & 1720$\pm$57 & 24.0 & (7) & 484  & (10)   & \FRI\  \\
0206+35	&          &0.0377	&159          & E         & -22.6  & $<$126      & $<$284      &	24.8 & (2) & 69.9 & (2,10) & \FRI\  \\
0222+36	&          &0.0334	&141          & E         & -22.2  & $<$126      & $<$315      &	23.7 & (4) & 4.8  & (5)    & C      \\
0258+35 & NGC 1167 & 0.0165     & 70          & S0        & -21.7  & 177$\pm$47  & $<$441      &	24.0 & (4) & 1.4  & (12)   & C      \\
0326+39	&          &0.0243	&103          & E$^{\dag}$ & -21.9    & $<$140      & $<$410      &	24.2 & (7) & 202  & (10)   & \FRI\  \\
0331+39	&          &0.0206	&87           & E         & -22.4    & $<$140      & $<$410      &	23.9 & (2) & 29.1 & (2,10) & \FRI\  \\
0648+27	&          &0.0412	&174          & S0        & -23.2    & 2758$\pm$57 & 2419$\pm$57 &	23.7 & (4) & 1.3  & (14)   & C      \\  
0722+30 &          & 0.0189     & 80          & S         & -20.5    & 3190$\pm$21 & 5141$\pm$57 &	23.0 & (4) & 13.6 & (2,10) & \FRI\      \\
0924+30	&          &0.0253	&107          & E$^{\dag}$ & -21.9   & $<$126      & $<$315      &	23.8 & (7) & 435  & (10)   & \FRI\  \\
1040+31	&          &0.0360	&152          & DB$^{\dag}$& -21.5   & $<$195      & $<$473      &	24.3 & (2) & 40.3 & (2,10) & \FRI\  \\
1108+27	& NGC 3563 &0.0331	&140          & S0        & -21.1$^{\dag}$ & $<$153$^{\ddag}$  & $<$315$^{\ddag}$ & 23.3 & (2) & 1085 & (10)   & \FRI\  \\
1122+39	& NGC 3665 &0.0069	&29           & S0        & -22.8    & 1813$\pm$47 & 7014$\pm$116&	22.0 & (5) & 10.7 & (2,10) & C      \\
1217+29	& NGC 4278 &0.0022	&16.1$^{1}$   & E         & -21.4     & 618$\pm$21  & 2041$\pm$57 &	22.2 & (6) & 0.009& (13)   & C      \\
1321+31	& NGC 5127 &0.0162	&68           & E pec     & -21.5    & $<$140      & $<$347      &	23.9 & (7) & 246  & (10)   & \FRI\  \\
1322+36 & NGC 5141 &0.0174	&73           & S0        & -21.5    & $<$153      & $<$378      &	23.7 & (2) & 19.1 & (2,10) & \FRI\  \\
1447+27	&          & 0.0306     & 129         & S0        & -21.4    & $<$112      & $<$284      &	23.6 & (6) & $<$2.3& (3)   & C      \\
1658+30 & 4C 30.31 & 0.0344     & 145         & E         & -20.9    & $<$112      & $<$252      &        24.2 & (3) & 114  & (5,10) & \FRI\  \\
2116+26 & NGC 7052 &0.0156	&66           & E         & -21.9    & 538$\pm$17  & 1276$\pm$57 &	22.7 & (2) & 291  & (10)   & \FRI\  \\
2229+39	& 3C 449   &0.0171	&72           & E         & -21.6    & 186$\pm$42  & 1029$\pm$137&	24.4 & (9) & 462  & (10)   & \FRI\  \\
\vspace{-2mm}        &          &            &            &          &          &             &             &      &	   &      &        &        \\
1557+26	&          & 0.0442     & 187         & E         & -22.1   & $<$126      & $<$315      & 23.1 & (2) & $\sim$2&(2)   & C      \\
\multicolumn{1}{c}{-}&NGC 3894  & 0.0108 & 46 & E         & -22.4   & 140$\pm$59  & 480$\pm$158 &	23.0 & (6) & 1.6  & (15)    & C      \\
\end{tabular}
\flushleft
{Notes -- The distance D (col. 4) to the radio galaxy and the linear size of the radio source (LS -- col. 11) have been determined from the redshift ({\sl z} -- col. 2) and $H_{0}$ = 71 \kms\ Mpc$^{-1}$ (unless otherwise indicated). Redshifts and optical morphology are based on results from the NASA/IPAC Extragalactic Database (NED), unless otherwise indicated. $M_{V}$ and IRAS flux densities (columns 6 - 8) are from \citet[][and refs. therein]{imp93} and adjusted for $H_{0}$ = 71 \kms\ Mpc$^{-1}$ (unless otherwise indicated). $^{\dag}$Information taken from \citet{bur79}. $^{\ddag}$Data taken from \citet{imp90}. {\sl Refs:} {\sl 1.} \citet{ton01}, {\sl 2.} \citet{par86}, {\sl 3.} \citet{rui86}, {\sl 4.} \citet{fan86}, {\sl 5.} \citet{fan87}, {\sl 6.} \citet{whi92}, {\sl 7.} \citet{eke81}, {\sl 8.} \citet{sch83}, {\sl 9.} \citet{lai80}, {\sl 10.} \citet[][based on the data used in this paper]{emo06thesis}, {\sl 11.} \citet{mor09_NGC315}, {\sl 12.} \citet*{gir05a}, {\sl 13.} \citet{wil98}, {\sl 14.} \citet{mor03}, {\sl 15.}. \citet{tay98}, \citet{bri79}}
\end{table*} 

Another advantage of studying \HI\ in radio galaxies is that \HI\ can be traced in absorption against the bright radio continuum. It allows investigating the kinematics of \HI\ gas located in front of the radio source, all the way down to the very nuclear region. This provides important insight in the presence/absence of circum-nuclear discs and tori, or allows to look for direct evidence of AGN fuelling/feedback in the form of gas infall/outflow \citep[see e.g.][]{gor89,mor01,mor08,ver03,mor05}

In this paper we study the large-scale \HI\ properties of a {\sl complete} sample of nearby low-power radio galaxies and combine this with deep optical observations of the low-surface brightness stellar content of the \HI-rich objects. The main aim is to investigate the importance of gas-rich galaxy mergers/interactions among low-power radio galaxies. We compare the \HI\ properties of our sample of nearby radio galaxies with similar studies done on radio-quiet early-type galaxies \citep{oos07,oos09}. Our sample of nearby radio galaxies consists of low-power compact and \FRI\ radio sources. The current paper succeeds a first {\sl Letter} in this series \citep[][hereafter \paperI]{emo07}, in which some \HI\ results related to the low-power compact sources in this sample were already discussed. While the current paper will shortly revisit the results from \paperI, it will also give the overall results and details on the entire sample. Results of a sample of more powerful \FRII\ radio sources as well as a discussion of the role of \HI\ gas in the \FRI/\FRII\ dichotomy will be presented in a future paper.\\
\ \\
Throughout this paper we use $H_{0}$ = 71 \kms\ Mpc$^{-1}$. We calculate distances using the Hubble Law c$\cdot$$z$=H$_{0}$$\cdot$D (i.e., for simplicity we assume a single value for both the luminosity distance and angular distance, which is accurate to within a few percent at the redshifts of our sample sources).

\section{The sample}
\label{sec:sample}

Our initial sample of nearby low-power radio galaxies consisted of 23 sources from the B2-catalogue \citep[][flux density limit $S_{\rm 408MHz} \ga 0.2$ Jy]{col70} with redshifts up to {\sl z} = 0.041. This initial sample is complete, with the restriction that we left out BL-Lac objects as well as sources in dense cluster environments (since here the gas content of galaxies is severely influenced by environmental effects [e.g. \citet{cay94,sol01,chu09}] and we expect that merger signatures may be wiped out on relatively short time-scales). Because of observational constraints, two sources were excluded from our initial sample (B2~1317+33 and B2~1422+26), but we do not expect that this will significantly alter our main results. Of our remaining sample of 21 radio galaxies, six have a compact radio source, while fifteen have an extended \FRI\ radio source (with `compact' defined as not extending beyond the optical boundary of the host galaxy, typically $\la 10$ kpc in diameter). Most of the compact sources have often been referred to as Low Power Compact (LPC) sources. The exception is B2~0258+35, which has been classified as a Compact Steep Spectrum (CSS) source \citep{san95}. In order to increase the number of compact sources in our sample, we observed two more radio galaxies with a compact source: B2~1557+26, a radio galaxy from the B2 catalogue with {\sl z} = 0.0442 (therefore just outside the redshift range of our complete sample) and NGC 3894 (a compact radio source that is comparable in power to our B2 sample sources, but with a declination outside the completeness limit of the B2 sample). While these two sources provide additional information on the \HI\ content of nearby radio galaxies with a compact source, they are left out of the statistical analysis of our complete sample discussed in the remainder of this paper. All the sources in our sample have a radio power of 22.0 $\leq$ log ($P_{\rm 1.4\ GHz}$) $\leq$ 24.8 and their host galaxies were a priori classified as early-type galaxies \citep[with the exception of the late-type system B2~0722+30;][]{emo09}. Table \ref{tab:sourceproperties} lists the properties of the radio galaxies in our sample.\footnote{In this paper we use the B2 name for both the radio source as well as the host galaxy.}

We note that our current sample contains no powerful \FRII\ radio galaxies. \FRII\ sources are generally located at a higher redshift and no \FRII\ source that meets our selection criteria is present in the B2 catalogue.

\begin{table}
\centering
\caption{Radio observations}
\vspace{0.4cm}
\label{tab:obsparam}
\begin{tabular}{llcc}
B2 Source & Observatory & Obs. date(s)     & t$_{\rm obs}$  \\
       &             & (dd/mm/yy)&(hrs) \\
\hline  
0034+25 & WSRT    & 09+11/08/07    & 24  \\
0055+30 & WSRT$^{a}$& 30/06/00;05/09/01& 21    \\
0104+32 & VLA-C   & 22/12/02    &  4   \\    
        & WSRT    & 23+29/08/07 & 24  \\
0206+35 & WSRT    & 29/08/04;22/08/07; & 31    \\
        &         & 11/09/07 & \\
0222+36 & WSRT    & 12+13/09/07 & 24   \\
0258+35 & WSRT$^{b}$& 22/10/06;12/07/08; & 107 \\
        &         & 19/08/08;26+29/09/08; &  \\
        &         & 01+02+07+10/10/08; \\
        &         & 07+12+17/11/08    \\
0326+39 & WSRT    & 17+18/09/07 & 24    \\ 
0331+39 & WSRT    & 19/08/04;19+20/09/07 & 30    \\
0648+27 & WSRT$^{c}$& 12+15/08/02;28/12/02  & 36 \\
0722+30 & VLA-C$^{d}$   & 23/12/02 & 4    \\
0924+30 & WSRT    & 07/08/04 & 9    \\
1040+31 & WSRT    & 31/07/04 & 7    \\ 
1108+27 & VLA-C   & 31/03/04 & 4    \\
1122+39 & VLA-C   & 20/04/04 & 2.5    \\  
1217+29 & WSRT$^{e}$& 04/02/04& 48 \\
1321+31 & VLA-C   & 02/11/02 & 4    \\
1322+36 & VLA-C   & 02/11/02 & 4    \\
1447+27 & WSRT    & 10/04/04 & 12   \\
1658+30 & WSRT   & 03+04/08/07 & 24  \\
2116+26 & WSRT    & 01+07/08/07 & 24   \\
2229+39 & WSRT    & 08+10/08/07 & 24   \\
        &         &          &      \\ 
1557+26 & WSRT    & 09/04/04 & 12   \\ 
NGC 3894& WSRT    & 01/02/04 & 12   \\
\end{tabular} 
\flushleft
{\sl Notes:} Although initially the WSRT and VLA observations were aimed at obtaining a uniform sensitivity, many of our sample sources were re-observed over the years, resulting in the varying observing times. t$_{\rm obs}$ (last column) is the total observing time. {{\sl References:} a). \citet{mor09_NGC315}; b). Struve et al. (in prep.) c). \citet{emo06}; d). \citet{emo09}; e). \citet{mor06b}}
\end{table}

\section{Observations}
\label{sec:observations}

\subsection{Neutral hydrogen gas}
\label{sec:obsHI}
 
Observations were done during various observing runs in the period Nov. 2002 - Nov. 2008 with the Very Large Array (VLA) in C-configuration and the Westerbork Synthesis Radio Telescope (WSRT). The C-configuration of the VLA was chosen to optimise the observations for sensitivity to detect both extended \HI\ emission as well as \HI\ absorption against the radio continuum and to match the beam of the WSRT (in order to obtain an as good as possible homogeneous \HI\ sample). For the WSRT observations we used the 20 MHz bandwidth with 1024 channels in two intermediate frequency (IF) modes. For the VLA-C observations we used the 6.25 MHz band with 64 channels and two IF modes. Table \ref{tab:obsparam} gives the details of the observations. 

For the reduction, analysis and visualisation of the data we used the {\tt MIRIAD}, {\tt GIPSY} and {\tt KARMA} software. After flagging, a standard bandpass-, phase- and (if necessary) self-calibration was performed on the data. In order to minimise aberration effects of strong continuum point-sources in the field of our target sources, the model components of these strong point sources were removed from the data in the uv-domain. Continuum and line data sets were constructed by fitting a first or second order polynomial to the line-free channels in the uv-data, applying a Fourier transformation and subsequently cleaning and restoring the signal in the data in order to remove the beam-pattern. The resulting continuum images do not have the optimal sensitivity and resolution for studying the radio sources in detail and are, therefore, omitted from this paper \citep[see][\ for a collection of the continuum images]{emo06thesis}. We note, however, that the radio source structures and flux densities in these images agree with continuum observations from the literature \citep[][]{par86,rui86,fan86,fan87}. For the line-data we constructed data cubes with different weighting-schemes in order to maximise our sensitivity for various emission/absorption features. Uniform weighting has been used to study in detail \HI\ in absorption against the radio continuum for most of our sample sources, while robust weighting \citep{bri95} provided the best results for tracing \HI\ in emission. Table \ref{tab:linedataproperties} gives an overview of the properties of the data sets that we used for this paper.

Total intensity maps of the line-data were made by summing all the signal that is present above (and below for absorption) a certain cut-off level in at least two consecutive channels. This cut-off level was determined at a few $\times$ the noise-level, the exact value depending on the noise properties of the individual data-cubes (but typically 3$\sigma$). In cases where the signal is very weak, it was taken into account only when it appeared in both polarisations and in both the first and the last half of the observations. Further details on the data reduction of several individual objects that we previously published can be found under the references mentioned in Table \ref{tab:obsparam}.

\begin{table*}
\centering
\caption{Properties of the \HI\ data}
\vspace{0.4cm}
\label{tab:linedataproperties}
\begin{tabular}{lccclcclc}
Source   &$\Delta$v & \multicolumn{3}{l}{{\bf Uniform -- absorption}}     & \multicolumn{4}{l}{{\bf Robust/Natural -- emission}}   \\
B2 name  & (km/s) & (1) & (2) & (3) & (1) & (2) & (3) & (4) \\
\hline  
0034+25        & 16.5 & 0.43 & $26.2 \times 11.8$ & (-1.0)  & 0.22  & $44.6 \times 25.6$ & (-2.7)& +2   \\ 
0055+30$^{a}$  & 20   & -    &           -        &   -     & 0.21  & $35   \times 18$   & (-5)   &+0.5  \\ 
0104+32{\small \ (VLA)}   & 20.6 & 0.27 & $13.4 \times 10.8$ & (-39.2) & 0.20  & $18.8 \times 18.1$ & (-40.5)& +2   \\
0104+32{\small \ (WSRT)}  & 16.5 & 0.41 & $22.2 \times 14.3$ & (17.2)  & 0.20  & $40.0 \times 26.7$ & (3.4)  & +2   \\
0206+35        & 16.5 & 0.34 & $21.4 \times 11.9$  & (9.2)  & 0.16  & $42.2 \times 24.7$ & (8.6)  & +2   \\
0222+36        & 16.5 & 0.43 & $16.7 \times 15.6$ & (39.2)  & 0.22  & $35.0 \times 30.4$ & (23.4) & +2   \\
0258+35$^{b}$  & 16.5 &   -  & -                  & -       & 0.13   & $28.9 \times 16.8$ & (1.5)  & +0.4   \\ 
0326+39        & 16.5 & 0.39 & $19.2 \times 13.4$ & (2.4)   & 0.19  & $38.1 \times 26.8$ & (-1.2) & +2   \\ 
0331+39        & 16.5 & 0.37 & $17.5 \times 14.7$ & (-11.1) & 0.18  & $38.7 \times 24.9$ & (1.3)  & +2   \\ 
0648+27$^{c}$   & 16.5 & 0.41 & $25.5 \times 11.2$ & (-0.4)  & 0.14  & $48.1 \times 24.2$ & (1.0)  & +1   \\
0722+30$^{d}$  & 20.6 & 0.29 & $11.9 \times 10.6$ & (2.8)   & 0.19  & $19.0 \times 17.5$ & (-24.5)& +2   \\
0924+30        & 16.5 & 0.51 & $25.1 \times 8.6$  & (10.2)  & 0.42  & $50.7 \times 19.0$ & (10.2) & +1   \\
1040+31        & 16.5 & 1.18 & $29.8 \times 9.7$  & (19.0)  & 0.52  & $60.5 \times 24.1$ & (20.7) & +2   \\  
1108+27        & 20.6 & 0.47 & $13.4 \times 11.1$ & (-41.8) & 0.31  & $19.5 \times 19.0$ & (88.8) & +2   \\
1122+39        & 20.6 & 0.79 & $15.0 \times 10.5$ & (-62.8) & 0.50  & $21.7 \times 17.7$ & (-81.3)& +2   \\  
1217+29$^{e}$  & 16.5 &  -   & -                  &   -     & 0.37  & $28 \times 14$   &  (11)    &  0   \\
1321+31        & 20.6 & 0.36 & $11.5 \times 11.3$ & (-25.6) & 0.28  & $16.5 \times 16.0$ & (-57.3)& +2   \\
1322+36        & 20.6 & 0.30 & $13.3 \times 12.5$ & (-72.8) & 0.26  & $15.1 \times 14.3$ & (-86.6)&+0.5  \\
1447+27        & 16.5 & 0.70 & $34.4 \times 13.6$ & (-1.9)  & 0.41  & $61.6 \times 24.6$ & (-0.9) & +2   \\
1658+30        & 16.5 & 0.39 & $23.8 \times 12.6$ & (1.9)   & 0.20  & $41.8 \times 24.6$ & (1.3)  & +2   \\
2116+26        & 16.5 & 0.38 & $25.2 \times 12.6$ & (-3.4)  & 0.19  & $45.3 \times 24.6$ & (-1.7) & +2   \\
2229+39        & 16.5 & 0.38 & $18.7 \times 13.2$ & (1.5)   & 0.19  & $35.6 \times 25.9$ & (-2.4) & +2   \\
               &      &      &                    &         &       &   	              &   &           \\  
1557+26        & 16.5 & 0.64 & $23.9 \times 11.8 $& (-0.3)  & 0.39  & $43.9 \times 20.7$ & (-0.1) & +1   \\
NGC~3894       & 8.2  & 0.71 & $13.2 \times 11.5$ & (0.4)   & 0.48  & $46.1 \times 41.3$ & (0.0)  & +2   \\
\end{tabular}\\
\flushleft 
{Notes -- $\Delta$v = channel separation; (1) = noise level (mJy beam$^{-1}$); (2) = beam-size (arcsec$^{2}$); (3) = position angle ($^{\circ}$); (4) robustness parameter. {\sl References:} a). \citet{mor09_NGC315}; b). Struve et al. (in prep.); c). \citet{emo06}; d). \citet{emo09}; e).\citet{mor06b}}
\end{table*}

\subsection{Optical imaging}
\label{sec:obsOpt}

Deep optical B- and V-band images were taken for all the radio galaxies in our sample with large-scale \HI\ gas detected in emission. The observations of all but one of our objects were done on 12, 13 and 14 March 2007 at the Hiltner 2.4m telescope of the Michigan-Dartmouth-MIT (MDM) observatory, located at the southwestern ridge of Kitt Peak, Arizona (USA). Imaging was done using the Echelle CCD, resulting in a field-of-view (F.o.V.) of $9.5 \times 9.5$ armcin. B2~0258+35 was observed on 15 November 2006 at the Hiltner 1.3m MDM telescope with the Templeton CCD, resulting in a F.o.V. of $8.5 \times 8.5$ arcmin. All observations were taken in relatively good to moderate seeing ($1-2$ arcsec) and under photometric conditions. Table \ref{tab:optparam} summarises the observational parameters. Because we are interested in studying very faint stellar features, detection of those features in both B- and V-band data assures their validity. In this paper we present the available B-band imaging, although we note that all the features presented and discussed in this paper are also detected in our V-band data.\footnote{For B2~0648+27 we present the co-added B+V-band data, as described in \citet[][]{emo08_0648}.}

\begin{table}
\centering
\caption{Optical observations}
\vspace{0.4cm}
\label{tab:optparam}
\begin{tabular}{llcccc}
B2 Source & MDM  & Obs. date & Int. time & airmass \\
\hline
0258+35   & 1.3m & 15/11/06 & 60 min& 1.0-1.1 \\
0648+27   & 2.4m & 13/03/07 & 60 min& 1.0-1.1 \\
0722+30   & 2.4m & 13/03/07 & 60 min& 1.0-1.2 \\
1217+29   & 2.4m & 14/03/07 & 40 min& 1.1-1.2 \\
1322+36   & 2.4m & 14/03/07 & 45 min& 1.1-1.3 \\
NGC~3894  & 2.4m & 14/03/07 & 50 min& 1.1     \\
\end{tabular} 
\end{table} 

We used the Image Reduction and Analysis Facility ({\tt IRAF}) to perform a standard data reduction (bias subtraction, flat-fielding, frame alignment and cosmic-ray removal). Probably due to minor shutter issues, a gradient was present in the background of the $9.5 \times 9.5$ arcmin CCD images obtained with the Echelle CCD. We were able to remove this effect to a significant degree by fitting a gradient to the background in the region surrounding our targeted objects and subsequently subtracting this background-gradient from our data. This method worked better for galaxies that covered only a small part of the CCD's F.o.V. (B2~0648+27 and B2~0722+30) than for galaxies that covered a large fraction of the CCD (B2~1217+29, NGC~3894 and B2~1322+36). The residual errors in the background subtraction are still visible in Fig. \ref{fig:HIsample}. This background issue made it impossible to obtain reliable flux and colour information from the B- and V-band images (in particular in the low-surface brightness regions). We therefore did not attempt an absolute or relative flux calibration of our sources. Using KARMA, we applied a world coordinate system to the images by identifying a few dozen of the foreground stars in a Sloan Digital Sky Survey (SDSS) image of the same region. The newly applied coordinate system agrees with that of the SDSS image to within 1 arcsec. This is good enough for comparing the optical with the \HI\ data, since the latter have a much lower resolution (see Table \ref{tab:linedataproperties}).

\section{Results}
\label{sec:results}

As can be seen in Table \ref{tab:hiproperties}, \HI\ in emission is associated with seven of the 23 radio galaxies in our sample, while nine sources show indications for \HI\ absorption against the radio continuum. 

Total intensity images of the \HI\ emission-line structures are shown in Fig. \ref{fig:HIsample}, together with deep optical imaging of their host galaxies \citep[B2~0055+30 is presented in][and therefore not repeated in Fig. \ref{fig:HIsample}]{mor09_NGC315}. Table \ref{tab:hiradiogalaxies} summarises the \HI\ properties. For five of the seven detections (B2~0258+35, B2 0648+27, B2 0722+30, B2 1217+29 and NGC 3894) the \HI\ gas is distributed in a regularly rotating disc- or ring-like structure with a mass of a few $\times 10^{8} - 10^{10} M_{\odot}$ and a diameter of several tens to hundreds of kpc. For two radio galaxies (B2~0055+30 and B2~1322+36), patchy \HI\ emission is observed, but the total mass associated with it is comparable to the upper limits that we derive for the non-detections. Since B2~0055+30 and B2~1322+36 are in the middle part of the redshift range of our sample sources, it is thus possible that sensitivity issues limit finding similar patchy, low-mass \HI\ emission in the higher redshift sources in our sample.

\begin{table*}
\centering
\caption{\HI\ emission and absorption results}
\vspace{0.4cm}
\label{tab:hiproperties}
\begin{tabular}{l|ccc|ccc}
Source & \HI\  & \HI\ mass & \HI\ contours & \HI\ & $\tau$ & $N_{\rm HI}$ ($T_{\rm spin} = 100$K) \\
B2 name& emission & ($\times 10^{8} M_{\odot})$ & ($\times 10^{20}$ cm$^{-2}$) & absorption & ($\%$) & $\times 10^{20}$ cm$^{-2}$ \\ 
\hline  
0034+25          & -          & $<$1.6  & -                                           & - & $<$8.7 & $<$16  \\   
0055+30$^{a}$     & +       &  0.66   & -                                           & + &   \multicolumn{2}{l}{\ \ \ \ 1\ \ \ \ \ (broad) \ \ \ \ 2.5} \\  	   	     
                    &   &         & -                                           &   &   \multicolumn{2}{l}{\ \ \ \ 5\ \ \ \ \  (narrow)\ \ \ 4.5} \\  
0104+32             & -          & $<$0.41 & -                                           & - & $<$0.3 & $<$0.6 \\   
0206+35             & -          & $<$1.6  & -                                           & - & $<$0.3 & $<$0.5 \\   
0222+36             & -          & $<$1.8  & -                                           & (+) & 1.0    & 1.0    \\   
0258+35$^{b}$       & +          & 180     & 0.26, 0.77, 1.3, 1.8, 2.3, 2.8, 3.3, 2.9, 4.4 & + &   0.23 & 1.2    \\   
0326+39             & -          & $<$0.82 & -                                           & - & $<$1.5 & $<$2.8 \\   
0331+39             & -          & $<$0.55 & -                                           & - & $<$0.2 & $<$0.3 \\   
0648+27$^{c}$       & +          & 85      & 0.22, 0.36, 0.52, 0.71,                     & + &   0.74 & 2.8    \\   
                    &            &         & 0.95, 1.2, 1.5, 1.8, 2.1                    &   &        &        \\   
0722+30$^{d}$       & +          & 2.3     & 0.63, 1.2, 1.8, 2.7, 3.3, 4.3, 5.3, 6.3, 7.7& + &    6.4 & 29     \\   
0924+30             & -          & $<$2.0  & -                                           & - & $<$38  & $<$69  \\   
1040+31$^{\dagger}$  & -          & $<$4.9  & -                                           & - & $<$1.1 & $<$2.0 \\   
1108+27             & -          & $<$2.8  & -                                           & - & $<$2.0 & $<$3.7 \\   
1122+39             & -          & $<$0.19 & -                                           & - & $<$11  & $<$20  \\   
1217+29$^{e}$       & +          & 6.9     & 0.1, 0.25, 0.5, 1.0, 2.5                    & - & $<$0.15$^{\ddagger}$ & $<$0.27       \\      
1321+31             & -          & $<$0.59 & -                                           & (+) &  \multicolumn{2}{l}{\ \ \ 5.5\ \ \ (nuc.)\ \ \ \ \ \ \ \ 4.9} \\   
                    &            &         & -                                           &   &  \multicolumn{2}{l}{\ \ \ 25\ \ \ \ (lobe)\ \ \ \ \ \ \ \ 36} \\   
1322+36             & +          & 0.69    & 1.7, 2.3, 2.8                               & + &    1.3 & 3.0    \\   
1447+27             & -          & $<$2.8  & -                                           & (+) &   0.87 & 2.9    \\   
1658+30             & -          & $<1.7$  & -                                           & - & $<$1.3 & $<$2.3 \\
2116+26             & -          & $<$0.34 & -                                           & - & $<$1.3 & $<$2.4 \\   
2229+39             & -          & $<$0.40 & -                                           & - & $<$0.7 & $<$1.2 \\   
                    &            &         & -                                           &   &        &        \\      
1557+26             & -          & $<$5.5  & -                                           & - & $<$6.2 & $<$11  \\
NGC~3894            & +          & 22      & 0.17, 0.49, 0.87, 1.7, 3.2, 4.6             & + &    4.1 &    14  \\
\end{tabular} 						   
\flushleft						
{`+' = detection, `(+)' = tentative detection, `-' = non-detection; Column 4 lists the \HI\ contours as shown in Figure \ref{fig:HIsample}; $^{\dagger}$We note that the \HI\ data of B2 1040+31 (taken with WSRT during service time) are of poor quality. Given the peculiar radio continuum morphology of B2 1040+31 \citep{par86}, this system deserves further \HI\ follow-up; $^{\ddagger}$Possible confusion with \HI\ emission; {\sl References:} a). \citet{mor09_NGC315}; b). Struve et al. (in prep.); c). \citet{emo06}; d). \citet{emo09}; e).\citet{mor06b}}
\end{table*}

\begin{figure*}
\centering
\includegraphics[width=\textwidth]{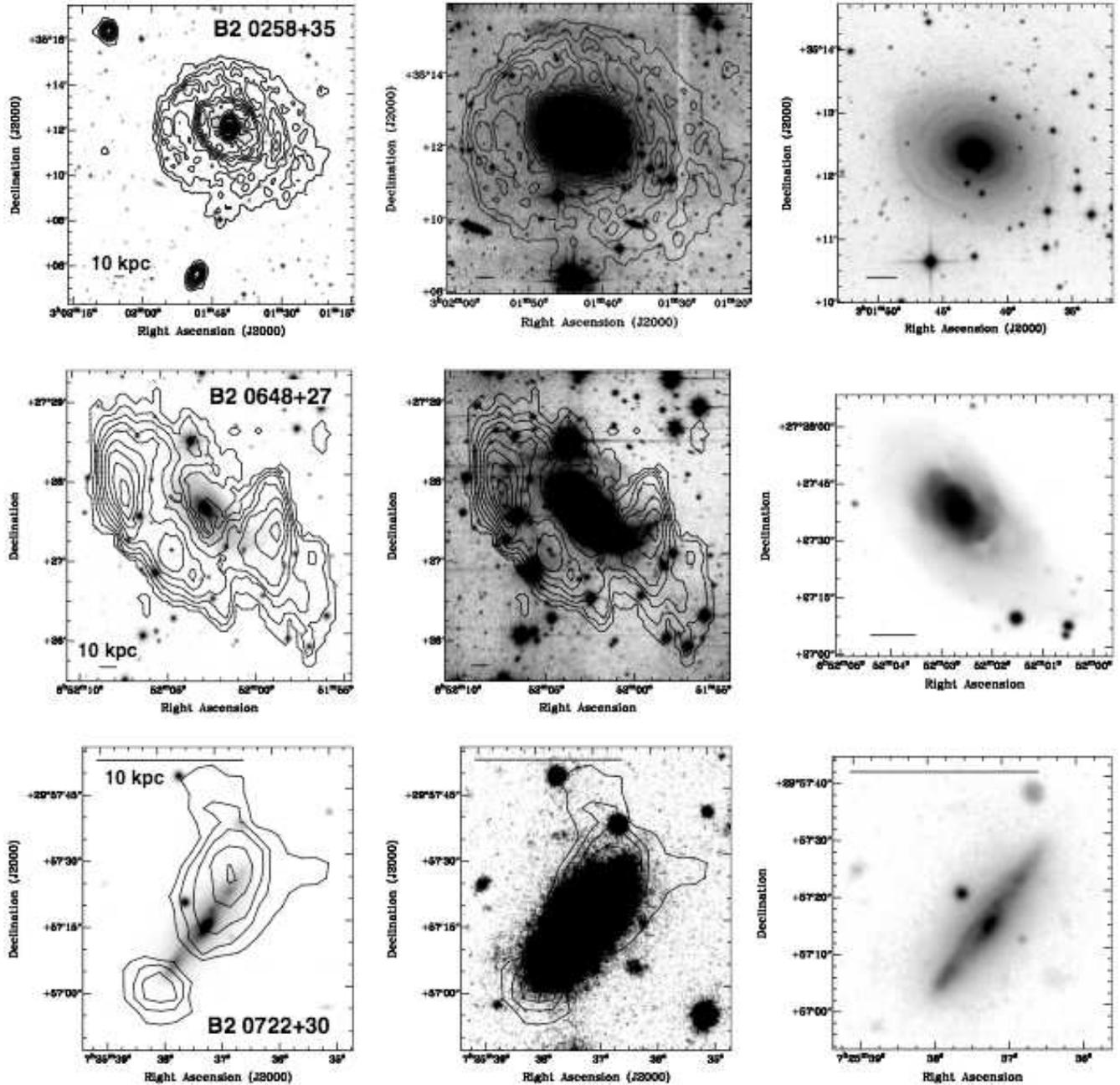}
\caption{Total intensity maps of \HI\ emission and deep optical imaging of our \HI-detected sample sources. Three plots are shown for each source; the left plot shows the \HI\ contours overlaid onto our deep optical image, the middle shows the same \HI\ contours overlaid onto a high-contrast representation of the deep optical image, the right plot emphasises the optical features of the main stellar body of the galaxy from our deep imaging. Contour levels of \HI\ emission are given in Table \ref{tab:hiproperties}.}
\label{fig:HIsample}
\addtocounter{figure}{-1}
\end{figure*}
\begin{figure*}
\centering
\includegraphics[width=\textwidth]{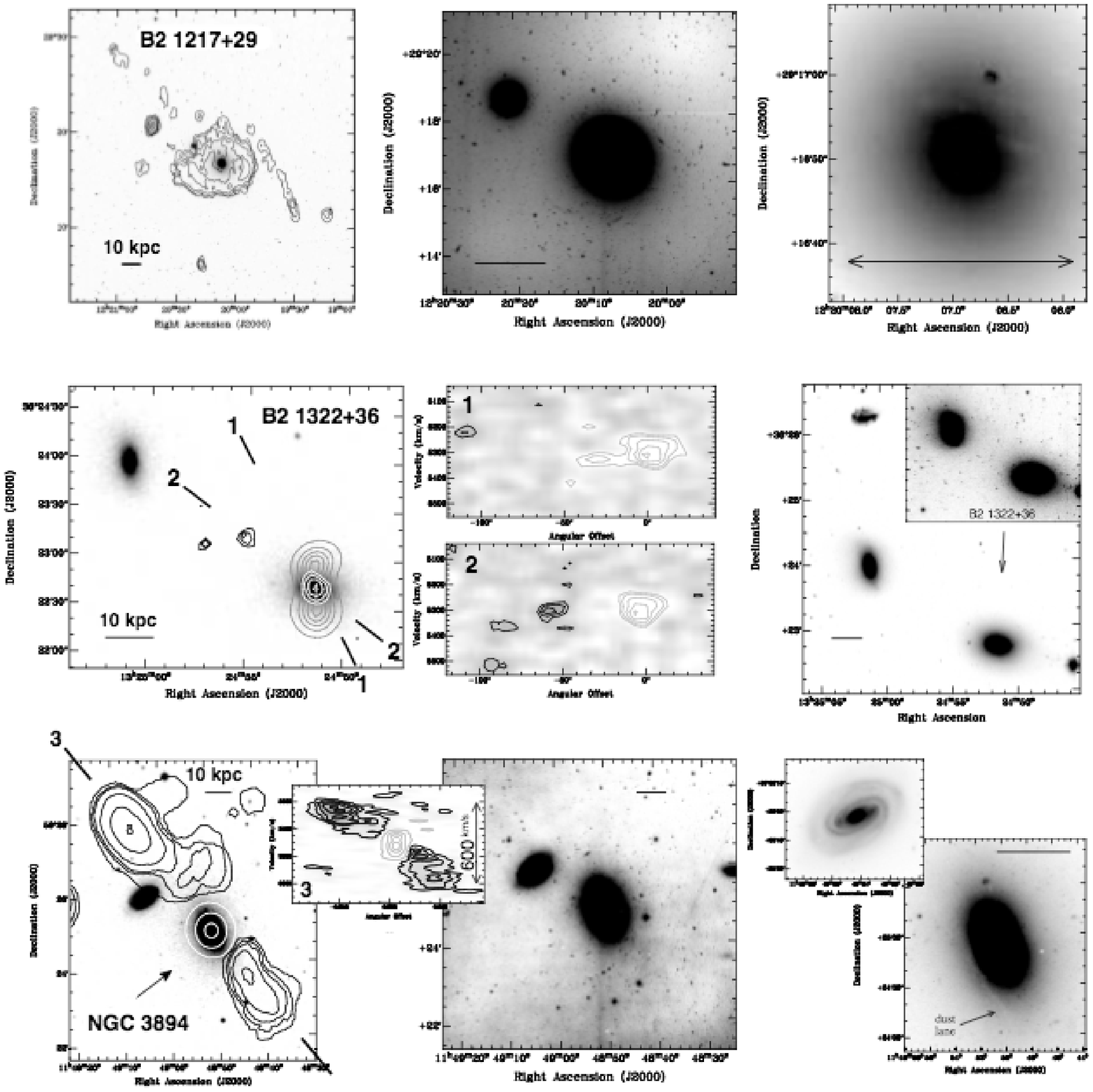}
\caption{{\sl -- continued.} The left plots show the contours of \HI\ emission overlaid onto our deep optical image (for B2~1322+36) or an optical SDSS image (for B2~1217+19 and NGC~3894), which has a larger field-of-view than our deep optical image. The \HI\ image of B2~1217+29 is taken from \citet{mor06b}. The left plot of B2~1322+36 also shows the contours of radio continuum (grey) and \HI\ absorption (white). For B2~1322+36 and NGC~3894, position-velocity (PV) plots of \HI\ emission (black contours) and absorption (grey contours) are also shown, taken along the lines indicated in the left plots. The middle plots of B2~1217+29 and NGC~3894 show a high-contrast representation of our deep optical image. The right plots emphasise the optical features of the main stellar body of the host galaxies (in particular the faint dust lanes in B2~1217+29 and NGC~3894) as well as several companion systems from our deep imaging. For all plots, contour levels of \HI\ emission are given in Table \ref{tab:hiproperties}. See \citet[][]{emo07} for more details on contour levels of \HI\ absorption, radio continuum and the PV-plots of B2~1322+36 and NGC~3894.}
\end{figure*}

\begin{figure*}
\centering
\includegraphics[width=\textwidth]{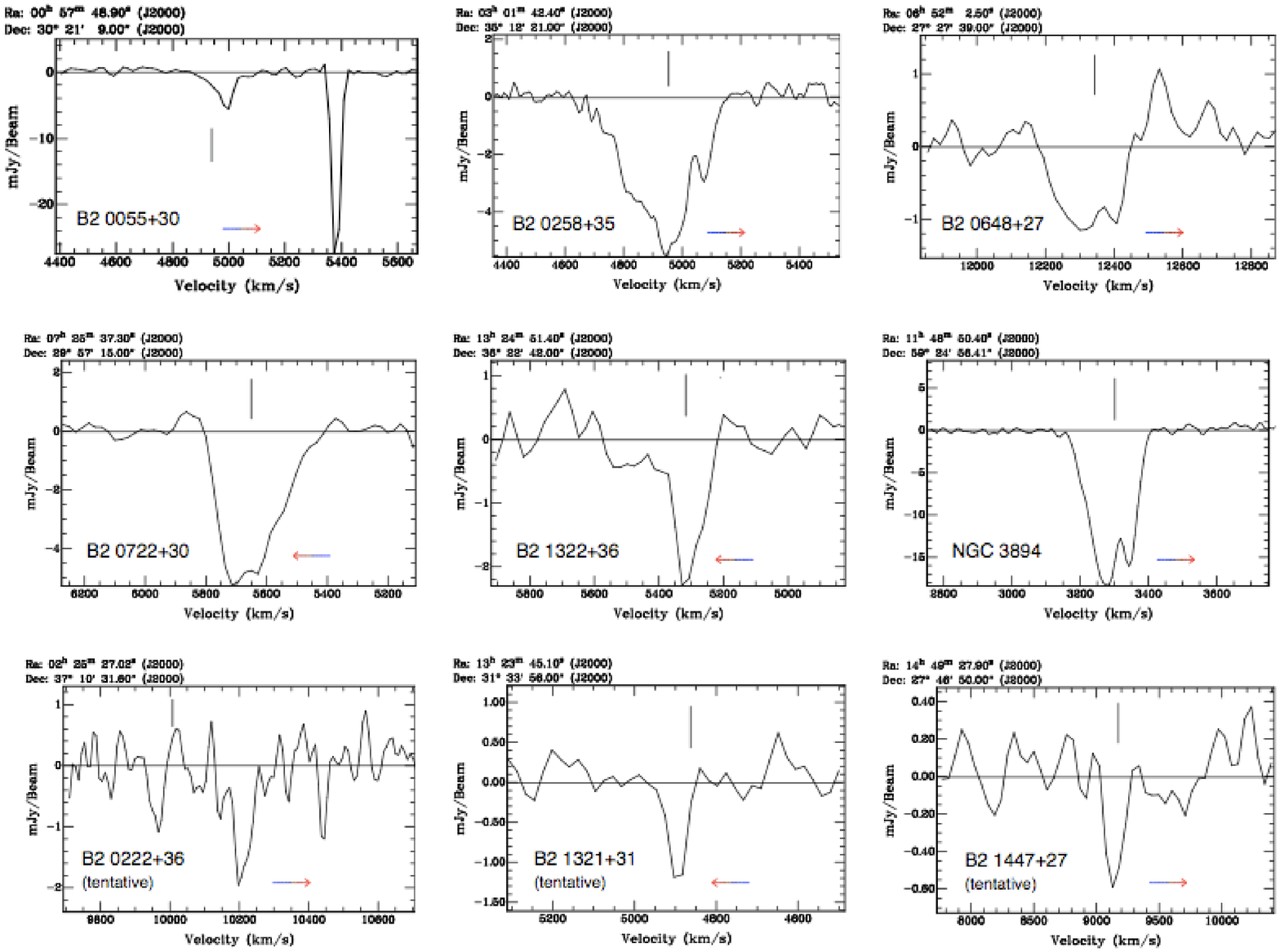}
\caption{Central \HI\ absorption profiles of our sample sources. The plot of B2~0055+30 is taken from \citet{mor09_NGC315}, while the plot of B2~0258+35 is from Struve et al. (in prep.). The velocities are given in optical definition. The bar indicates the systemic velocity traced with optical emission lines. Values of v$_{\rm sys}$ are taken from NED (unless otherwise indicated in Appendix \ref{app:hiproperties}). The arrow indicates the direction of increasing redshift velocity -- the right and left pointing arrows correspond to the WSRT and VLA data respectively. Our classification of `tentative' is based on the weakness of the `signal' in combination with the quality of the data-cubes.}
\label{fig:absorption}
\end{figure*}

\begin{figure*}
\centering
\includegraphics[width=0.7\textwidth]{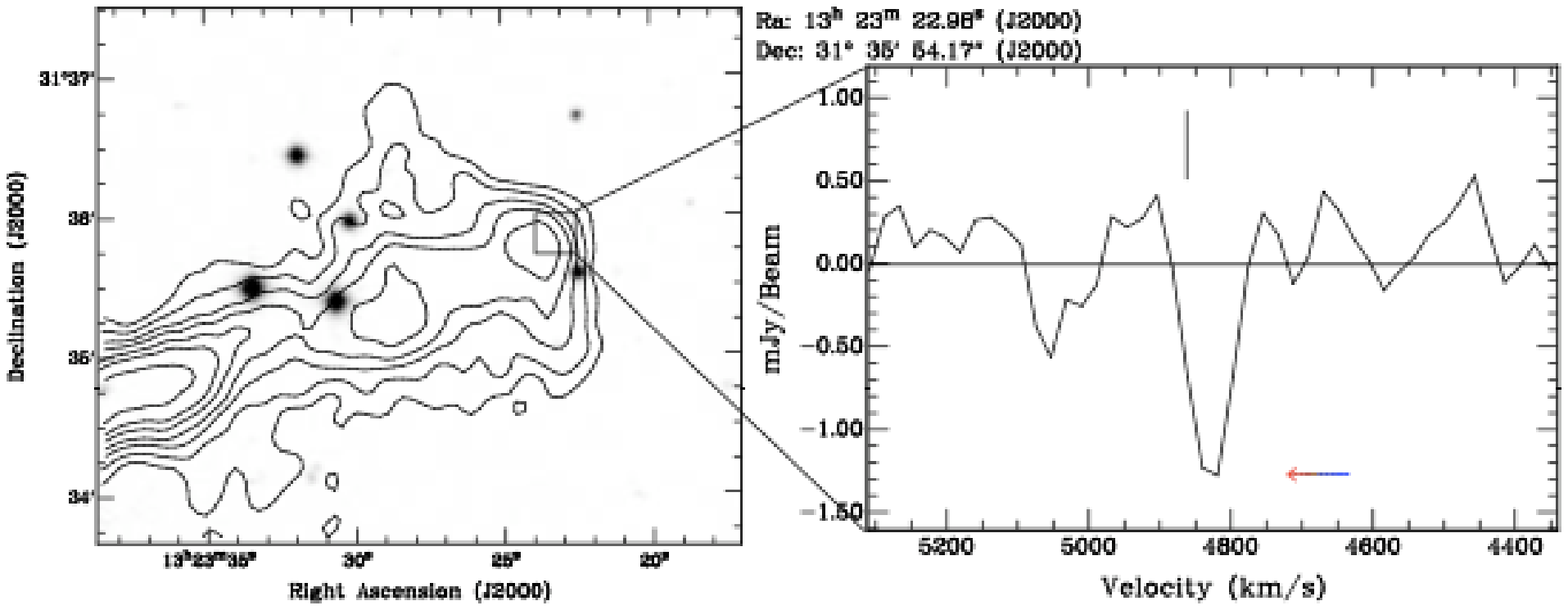}
\caption{Tentative \HI\ absorption profile of B2 1321+31 against the outer western radio lobe.}
\label{fig:absorption1321_extended}
\end{figure*}

\begin{figure*}
\centering
\includegraphics[width=\textwidth]{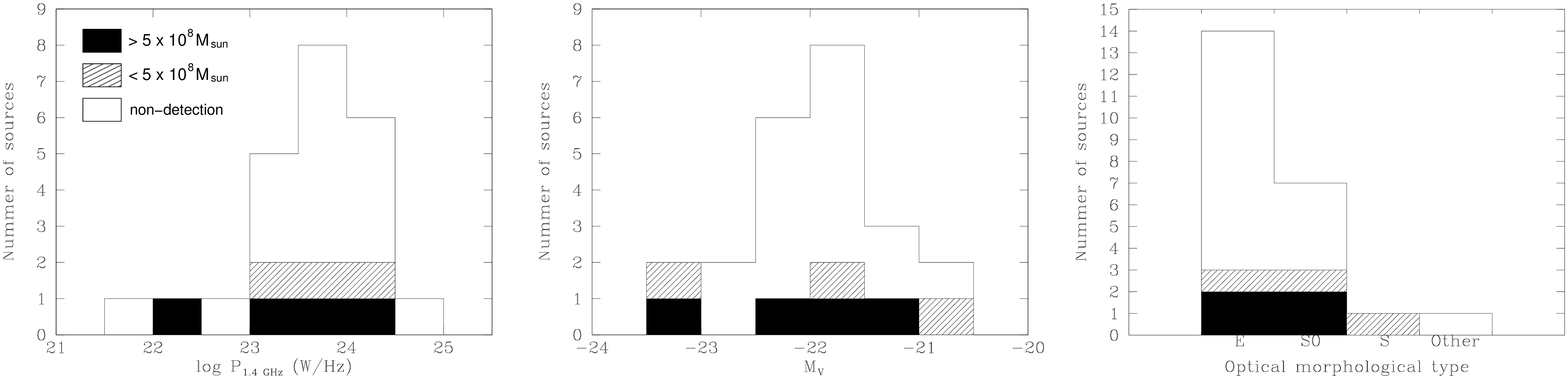}
\caption{Histogram of the \HI-emission detections and non-detections in our sample regarding the total power of the radio source at 1.4 GHz {\sl (left)} as well as $M_{\rm V}$ {\sl (middle)} and morphological type {\sl (right)} of the host galaxy. (Values are taken from Table \ref{tab:sourceproperties}.)}
\label{fig:histograms}
\end{figure*}

The upper limits for the non-detections are estimated assuming a potential 3$\sigma$ detection smoothed across a velocity range of 200 \kms\ (therefore resembling the large-scale \HI\ structures that we detect):

\begin{equation}
\frac{M_{\rm upper}}{M_{\odot}} = 2.36 \cdot 10^{5} \times D^{2} \times S_{3\sigma} \times \Delta v \times \sqrt{\frac{200\ \rm {km/s}}{\Delta v}},
\label{eqn:3_upper}
\end{equation}
where $S_{3\sigma}$ is the 3$\sigma$ noise level per channel (in Jy beam$^{-1}$, from the robust weighted data), $D$ the distance to the galaxy (in Mpc) and $\Delta$v the channel width (in \kms)  - see Tables \ref{tab:sourceproperties} and \ref{tab:linedataproperties}.

H\,{\sc i} absorption is unambiguously detected against the radio continuum of six of our sample sources, while another three show tentative evidence for absorption, which has to be confirmed with additional observations (see Fig. \ref{fig:absorption}). All sources for which \HI\ has been detected in emission also show unambiguous \HI\ absorption, except B2~1217+29. For all nine sources that show (tentative) absorption, an \HI\ absorption profile is seen against the central region of the galaxy. For eight of the nine sources the central absorption is spatially unresolved. Only for B2 1322+36 the absorption is slightly extended against the resolved radio continuum (see Fig. \ref{fig:HIsample}). Two sources (B2~0055+30 and B2~1321+31) show evidence for multiple absorption components. B2~0055+30 shows two components against the central radio continuum \citep[a broad and a narrow one, see][\ and Appendix \ref{app:hiproperties}]{mor09_NGC315}, while B2~1321+31 shows a second, spatially unresolved, tentative component against the outer edge of one of the radio lobes (see Fig. \ref{fig:absorption1321_extended}). Table \ref{tab:hiproperties} includes both components for these two sources.

Table \ref{tab:hiproperties} lists the optical depth and column density of the absorption features. The optical depth ($\tau$) is calculated from: 
\begin{equation}
e^{-\tau} = 1 - \frac{S_{\rm abs}}{S_{\rm cont}},
\label{eqn:3_tau}
\end{equation}
where $S_{\rm abs}$ is the peak flux density of the absorption and $S_{\rm cont}$ the flux density of the underlying radio continuum. Subsequently, the \HI\ column density ($N_{\rm HI}$) is given by:
\begin{equation}
N_{\rm HI}\ (\rm cm^{-2}) = 1.8216 \cdot 10^{18} \times T_{\rm spin} \times \int \tau(v)\ dv,
\label{eqn:3_NHI}
\end{equation}
where $v$ is the velocity and $T_{\rm spin}$ the typical spin temperature of the \HI\ gas, assumed to be 100K. The \HI\ column densities have been derived assuming a covering factor of 1 for the gas that overlies these radio sources. For non-detections an upper limit is calculated assuming a potential 3$\sigma$ detection (uniform weighting) spread over 100 \kms.\\
\vspace{0mm}\\
Appendix \ref{app:hiproperties} gives a detailed description of the individual objects in which \HI\ is detected in emission or absorption. In the remainder of this Section we will describe the general \HI\ properties of the sample as a whole. Section \ref{sec:hiemission} will summarise the \HI\ emission properties, followed by the \HI\ absorption results in Sect. \ref{sec:hiabsorption}.\\

\subsection{H{\tiny \ }{\small I} emission}
\label{sec:hiemission}

\HI\ emission has been detected in 7 of the 23 sample galaxies. When taking into account only the complete sample (so without B2 1557+26 and NGC 3894; Sect. \ref{sec:sample}), our detection rate is 29$\%$. In Sect. \ref{sec:radioquiet} we compare this detection rate with that of radio-quiet early-type galaxies.

Before summarising the sample properties in detail, we first check whether the presence of large-scale \HI\ emission-line gas depends on some important parameters of the galaxies in our sample. Figure \ref{fig:histograms} shows histograms of the distribution of both the \HI\ detections and non-detections regarding the optical morphological class and absolute visual magnitude ($M_{\rm V}$) of the host galaxy, as well as the total power ($P_{\rm 1.4GHz}$) of the radio source. Although we have to be careful with our small-number statistics, there is no apparent bias in detecting \HI\ regarding these various observables (see Sect. \ref{sec:hiemission_optical} for a more in-depth discussion on the optical morphologies of the host galaxies). It is interesting to note that the range of absolute visual magnitudes among our sample sources covers the intermediate- and high-mass end of the galaxies from the Sloan Digital Sky Survey used in the Colour-Magnitude relations by \citet{bal04}. We therefore do not observe a difference in \HI\ content among early-type galaxies of different mass in our sample.

Above an \HI\ mass of $10^{8} M_{\odot}$, all \HI\ structures detected in our sample are fairly regularly rotating discs or rings (although a varying degree of asymmetry is still visible in these structures). We find no clear evidence for {\sl ongoing} gas-rich mergers in the form of long gaseous tidal debris (although B2~0648+27 is clearly a post-merger system; see Appendix \ref{app:hiproperties}). One potential worry is that the sensitivity of our \HI\ observations is not ideal for detecting low surface brightness tidal features that are not (yet) settled. \citet*{gre04} show that for decreasing sensitivity to detect \HI\ in emission, complicated velocity structures in \HI\ tend to wash out and the \HI\ often gets a more smooth and rotating appearance. Nevertheless, in \citet{emo06thesis} we showed examples of galaxies within the F.o.V. of our radio sources (but physically unrelated) that do show extended and complex tidal \HI\ structures, which shows that we are sensitive enough for observing such features at the redshift of our sample sources.

Table \ref{tab:hiradiogalaxies} gives $M_{\rm H{\small\ I}}/L_{\rm V}$ for our \HI\ detected radio galaxies. The large spread in $M_{\rm H{\small\ I}}/L_{\rm V}$ for these galaxies (ranging from 0.0005 to 0.44 $M_{\odot}/L_{\odot}$) is consistent with a large spread of $M_{\rm H{\small\ I}}/L_{\rm B}$ found by \citet*{kna85} and \citet{mor06b} for elliptical galaxies (because our B-band data were not optimised for photometric studies, we had to rely on available V-band magnitudes from the literature; see Table \ref{tab:sourceproperties}). The large spread in $M_{\rm H{\small\ I}}/L_{\rm B}$ for elliptical galaxies compared to spiral galaxies has led \citet{kna85} and \citet{mor06b} to conclude that the \HI\ gas in ellipticals is decoupled from the stars and has an external origin. The possible formation mechanism of the large-scale \HI\ discs/rings in our sample sources will be discussed in detail in Sect. \ref{sec:HIrich}.

\begin{figure}
\centering
\includegraphics[width=0.46\textwidth]{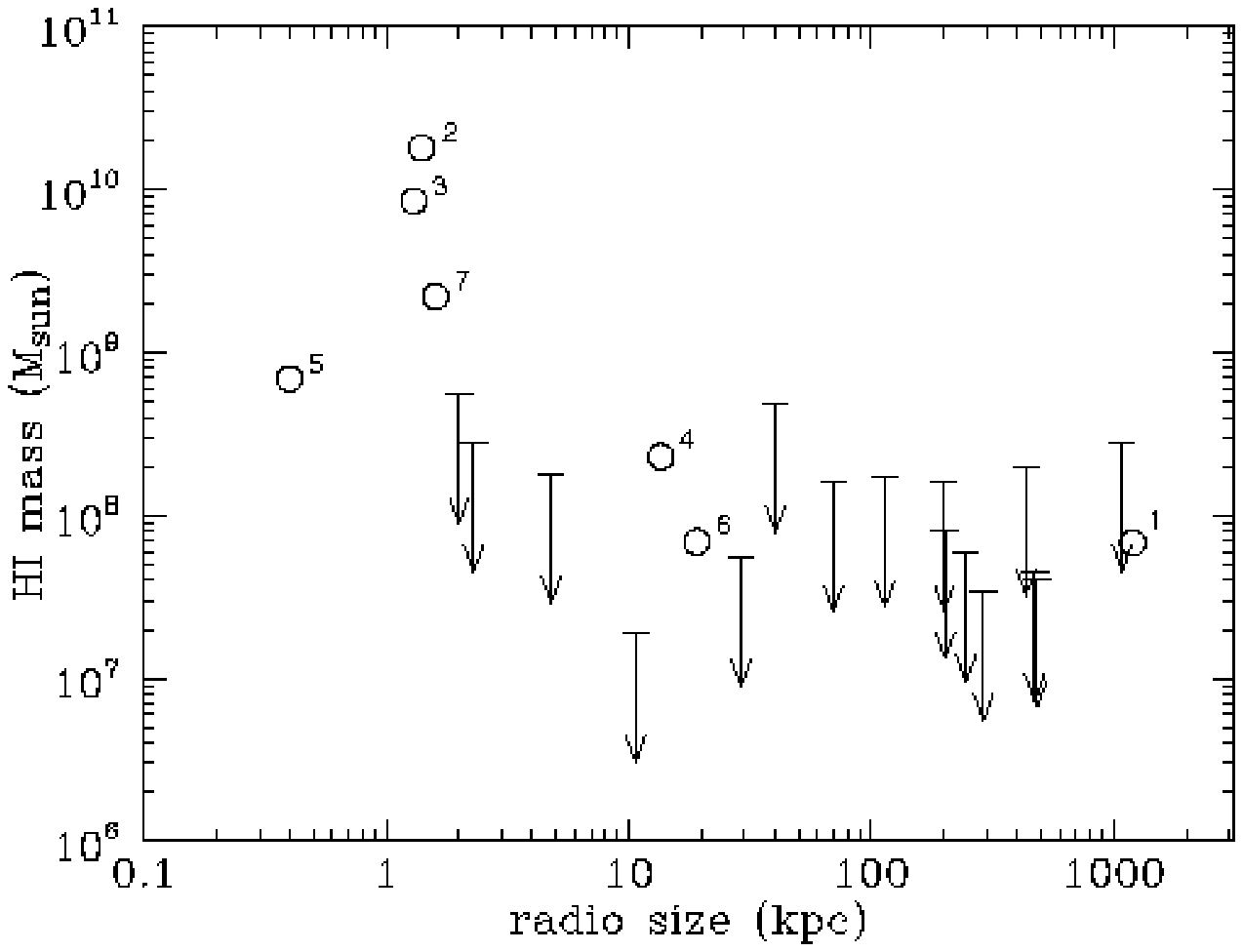}
\caption{Total \HI\ mass detected in emission plotted against the linear size (diameter) of the radio sources. In case of non-detection a tight upper limit (3$\sigma$ across 200 km s$^{-1}$) is given. The numbers correspond to the sources as they are given in Table \ref{tab:hiradiogalaxies}.}
\label{fig:HImasssize}
\end{figure}

Perhaps the most intriguing result from our \HI\ study is that galaxies with large amounts of extended \HI\ ($M_{\rm HI} \ga 10^{9} M_{\odot}$) all have a {\sl compact} radio source, while none of the host galaxies of the more extended FR-I type radio sources shows similar amounts of \HI. This is illustrated in Fig. \ref{fig:HImasssize}, where we plot the total mass of \HI\ detected in emission against the linear size of the radio sources. In \paperI\ we already presented and discussed this segregation in large-scale \HI\ content between compact sources and extended \FRI\ sources, suggesting that {\sl there is a physical link between the properties of the central radio source and the large-scale properties of the ISM}. In Sect. \ref{sec:relation} we will briefly review our conclusions from \paperI.

\begin{figure*}
\centering
\includegraphics[width=0.9\textwidth]{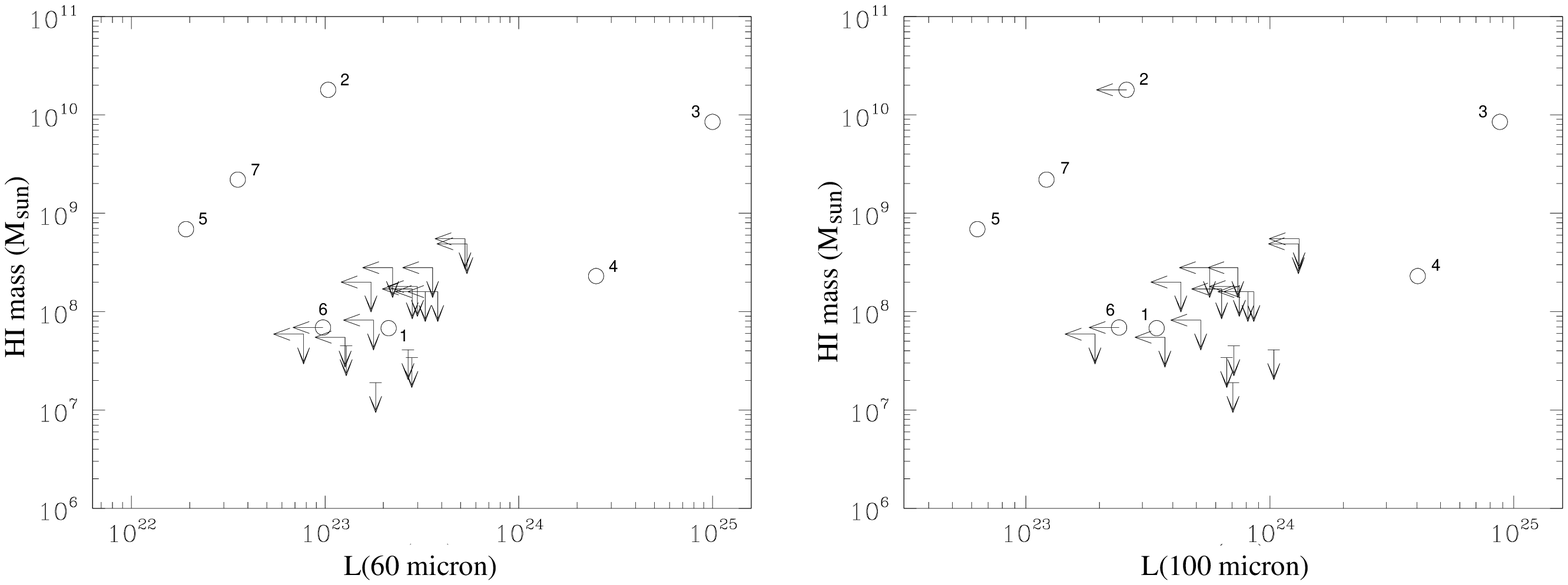}
\caption{Total \HI\ mass detected in emission plotted against the 60$\mu$m (left) and 100$\mu$m (right) IRAS luminosity. The arrows represent upper limits. The numbers correspond to the sources as they are given in Table \ref{tab:hiradiogalaxies}.}
\label{fig:iras}
\end{figure*}

All but one (B2~1322+36) of the radio sources in our sample that contain \HI\ emission are also detected in the infra-red (IR) at 60$\mu$m \citep[see Table \ref{tab:sourceproperties}; IR data are taken from][and references therein]{imp93}. However, as can be seen from Fig. \ref{fig:iras}, when taking into account the distance to the sample sources and hence converting the IR flux-density to IR luminosity (using the simple conversion $L_{\nu} = 4 \pi D^{2} S_{\nu}$), there is no clear correlation between large-scale \HI\ mass content and IR-luminosity. It is interesting, though, that B2~0648+27 and B2~0722+30 have by far the highest IR-luminosity of our sample sources (both at 60 and 100$\mu$m). The 60$\mu$m emission is expected to trace cool dust that could predominantly be heated by young stars \citep[e.g.][]{san96}. Indeed, spectral analysis revealed evidence for a prominent young stellar population throughout the host galaxy for both B2 0648+27 and B2 0722+30 \citep{emo06,emo09}. Based on the IR-luminosity alone, we do not expect young stellar populations as prominent as in B2~0648+27 and B2~0722+30 to be present in the other radio sources in our sample \citep[although a smaller contribution from young stars cannot be ruled out; see e.g.][]{wil04}.

\subsubsection{Optical morphology}
\label{sec:hiemission_optical}

Despite the morphological and kinematical similarity of the large-scale \HI\ discs/rings, the optical host galaxy morphology of the \HI\ detected radio galaxies varies significantly (see Fig. \ref{fig:HIsample}). While the merger remnant B2~0648+27 shows a distorted structure and a faint stellar ring, B2~0722+30 and B2~0258+25 contain a clear stellar disc. These three systems therefore contain a (faint) optical counterpart to the large-scale \HI\ structure. Contrary, B2~1217+29 and NGC~3894 have the apparent morphology of dust-lane ellipticals and only a bulge component is visible in our deep optical imaging. We note, however, that these two galaxies fill a substantial part of the CCD and serious limitations in the background subtraction of the optical data limit our ability to trace faint stellar light across the region where the \HI\ stretches (see Sect. \ref{sec:observations}). Moreover, NGC~3894 contains a very faint dust-lane that stretches in the same direction as the \HI\ disc. Since the \HI\ disc is viewed edge-on, it is possible that significant extinction may obscure a very faint stellar counterpart to the large-scale \HI\ disc.

\begin{table*}
\centering
\caption{{\sl HI around radio galaxies.} Given is the name, the NGC number, the total \HI\ mass detected in emission, the diameter of the \HI\ structure (or distance to the host galaxy for B2 1322+36), the peak in \HI\ surface density, the relative \HI\ content and the morphology of the \HI\ structure (D = disc, R = ring, C = cloud).}
\label{tab:hiradiogalaxies}
\begin{tabular}{llcccccc}
$\#$ & B2 Name & NGC & M$_{\rm HI}$ & D$_{\rm HI}$ & $\Sigma_{\rm HI}$ & $M_{\rm HI}/L_{\rm V}$ & Mor. \\
 &  &  & (M$_{\odot}$) & (kpc) &(M$_{\odot}$/pc$^{2}$) & ($M_{\odot}/L_{\odot})$ & \HI \\
\hline
\hline
1 & 0055+30$^{\ a}$ & 315   &  6.8$\times$10$^{7}$  & 10 & 1  & 0.0005 & C \\ 
2 & 0258+35$^{\ b}$ & 1167  &  1.8$\times$10$^{10}$ & 160 & 2.7 & 0.44   & D \\
3 & 0648+27$^{\ c}$ & -     &  8.5$\times$10$^{9}$  & 190 & 1.7 & 0.052  & R \\
4 & 0722+30$^{\ d}$         & -     &  2.3$\times$10$^{8}$  & 15  & 4.1 & 0.017  & D \\
5 & 1217+29$^{\ e}$ & 4278  &  6.9$\times$10$^{8}$  & 37  & 2.0 & 0.022  & D \\
6 & 1322+36         & 5141  &  6.9$\times$10$^{7}$  & 20  & 3.7 & 0.002 & C \\
7 & -               & 3894  &  2.2$\times$10$^{9}$  & 105 & 3.8 & 0.028  & R/D \\
\hline
\hline
\end{tabular}\\
\vspace{1mm} 
{\sl References:} a). \citet{mor09_NGC315}; b). Struve et al. (in prep.); c). \citet{emo06}; d). \citet{emo09}; e). \citet{mor06b}
\end{table*}

\subsubsection{H{\tiny \ }{\small I} environment}
\label{sec:environment}

Many of our sample sources contain \HI-rich galaxies in their environment. However, with the exception of the late-type galaxy B2 0722+30 \citep{emo09}, none of our targeted B2 radio galaxies shows any obvious evidence in \HI\ for ongoing interactions with \HI-rich companions (in the form of tidal-bridges, -arms or -tails). Features such as the clouds of \HI\ gas in between B2~1322+36 and its companion, the faint tails of \HI\ gas stretching off the disc in B2~1217+29 and the slight distortion in the \HI\ discs around B2~0258+35 and NGC~3894 could, however, present more subtle indications for less violent, gas-poorer or older interactions. A quantitative study of gas-rich companions in the environment of our sample sources would be interesting for estimating the \HI\ accretion rate/probability in nearby radio galaxies through such less violent galaxy encounters. However, this is beyond the scope of the current paper and will be presented in a future publication by Struve et al. (in prep).

\subsection{H{\tiny \ }{\small I} absorption}
\label{sec:hiabsorption}

H\,{\sc i} is unambiguously detected in absorption against the radio continuum for 6 of the 23 sample sources, while three more sources show a tentative detection (see Fig. \ref{fig:absorption}). The detection rate of \HI\ absorption in our complete sample is therefore $24-38 \%$ (depending on whether or not the three tentative detections are included). For the compact sources and extended \FRI\ sources, the detection rates are $33-67 \%$ and $20-27 \%$ respectively. 

Our detection rate of extended \FRI\ sources is slightly higher than that derived by \citet{mor01} (who detect \HI\ absorption in 10$\%$ of \FRI\ sources from the 2-Jy sample). However, when excluding the rare disc-dominated radio galaxy B2~0722+30 from our statistics (see Appendix \ref{app:hiproperties}), our detection rate drops to $14-21 \%$. This is in reasonable agreement with the values found by \citet{mor01}, given the low number statistics in their 2-Jy sample and the fact that their upper limits on the optical depth are almost a factor 2 larger than those in our B2 sample. It is interesting to note, however, that the \FRI\ radio sources in the 2-Jy sample of \citet{mor01} are on average more than an order of magnitude more powerful at 1.4 GHz than the \FRI\ sources in our B2 sample. These \HI\ absorption results therefore suggest that more powerful \FRI\ radio sources do {\sl not} have a higher detection rate of \HI\ absorption compared with less powerful \FRI\ sources. 

Our detection rate of compact sources is in good agreement with the detection rate of $54 \%$ that \citet*{pih03} derive for a large sample of Gigahertz Peaked Spectrum (GPS) and Compact Steep Spectrum (CSS) sources, despite the significantly lower radio power of our sources. Interestingly, \citet{pih03}, and later \citet[][]{gup06}, detect an anti-correlation between the projected linear size of compact radio sources and the \HI\ column density. Since this anti-correlation is attributed to a gradient in the distribution of the cool ISM in the central region of the radio galaxies, it seems that it is not immediately related to the segregation that we find in \HI\ content between compact and extended sources (Sect. \ref{sec:hiemission}). 

For most cases, the peak of the \HI\ absorption appears to coincide with the systemic velocity as derived from optical emission lines. For B2~0055+30, the \HI\ absorption is clearly redshifted with respect to the systemic velocity and part of it could represent gas falling into the nucleus \citep{mor09_NGC315}. For two other sources (B2~0222+36 and B2~1321+31) there are also indications that the peak of the \HI\ absorption is redshifted with respect to v$_{\rm sys}$, but these detections are only tentative and we argue that the uncertaintly in v$_{\rm sys}$ determined from optical emission-line is too large to make any claims. Many of the \HI\ absorption profiles in our sample are resolved in velocity and show both blue- and redshifted components, consistent with what is frequently observed in compact radio sources \citep{ver03,pih03,gup06}.

In contrast to the \HI\ emission results, there is no clear trend between radio source size and the presence of \HI\ absorption. We note, however, that the strength of the underlying radio continuum as well as the geometry of the absorbing \HI\ gas are important selection effects that influence our absorption results, while they are not relevant for detecting \HI\ in emission. 

All the galaxies in our sample that are unambiguously detected in absorption also show \HI\ emission-line structures at the same velocity. In fact, only one galaxy in the sample that is detected in emission is not detected in absorption, namely B2~1217+29, although \citet{mor06b} show that the large-scale \HI\ matches very well the ionised gas in the central region of this galaxy. The remaining three \HI\ absorption systems show only tentative detections. {\sl This strongly suggests that at least a significant fraction of the \HI\ gas in many nearby absorption systems is part of gaseous structures on scales larger than just the (circum)-nuclear region.} This is also in agreement with the idea that \FRI\ sources do not necessarily require a geometrically thick torus, as already suggested by \citet{mor01} from the above mentioned low detection rate of \HI\ absorption among \FRI\ sources in the 2-Jy sample \citep[although we note that there are \FRI\ sources for which the AGN is hidden by dust, e.g.][]{lei09}.

The \HI\ absorption characteristics of B2~0055+30 and B2~1322+36 indicate that extended \HI\ does occur in some \FRI\ radio galaxies, but generally in much lower amounts than that associated with a significant fraction of the compact sources in our sample. However, the low detection rate of off-nuclear \HI\ detected in absorption against the extended radio lobes of \FRI\ sources also suggests that such extended \HI\ structures are not a commonly observable feature among \FRI\ radio galaxies.

\section{Discussion}
\label{sec:discussion}

The \HI\ results presented in this paper provide -- for the first time -- a systematic insight into the properties of cold gas in nearby, low-power radio galaxies. In this Section we discuss these \HI\ results in order to investigate the nature of low-power radio galaxies in more detail. In Sect. \ref{sec:HIrich} we first summarise the \HI\ characteristics of our sample and conclude that, generally, we find no clear evidence for ongoing gas-rich galaxy mergers and interactions among our sample sources. Section \ref{sec:relation} discusses the fact that large amounts of \HI\ gas are only found around compact radio sources in our sample, while extended \FRI\ sources lack similar amounts of \HI\ gas. The possible explanations for this segregation in \HI\ mass with radio source size are revisited from \paperI. In Sect. \ref{sec:radioquiet} the \HI\ properties of our sample of low-power radio galaxies are compared with those of radio-quiet early-type galaxies, and from that comparison we find no clear differences. Section \ref{sec:nature} summarises our understanding of the nature of low-power radio galaxies.

\subsection{\HI\ characteristics}
\label{sec:HIrich}

Large-scale \HI\ is associated with seven of the radio galaxies in our sample. The overall \HI\ properties of our sample sources show a morphological trend in the sense that toward the high-mass end all the \HI\ structures are fairly regularly rotating discs/rings. Only the two \HI\ detections at the low-mass end (several $\times 10^{7} M_{\odot}$) appear much more irregular/clumpy. The regular kinematics of the large-scale \HI\ discs/rings in our sample suggest that the gas is either settled or in the process of settling. The morphology of these large-scale \HI\ structures is fairly uniform, although their sizes differ significantly (from 15 to 190 kpc) and their optical host galaxies show a range of morphologies. 

In this Section we discuss in detail what physical processes may have formed the observed \HI\ structures in our sample. This provides us with information about the evolutionary history of the host galaxy, which will be useful for the remainder of the Discussion.\\
\vspace{0mm}\\
{\sl Major merger or collision:}\\
We find no evidence for {\sl ongoing} gas-rich major mergers (i.e. mergers between galaxies with roughly equal mass) or massive galaxy collisions - in the form of large-scale tidal tails or bridges of \HI\ gas - among our sample sources. However, as we described in detail in \paperI, the formation of a large-scale disc-like structure may be the natural outcome of a major merger between gas-rich galaxies over the time-scale of one to several Gyr \citep[which at the same time also results in the formation of an early-type host galaxy from the merging systems;][]{hib96}. This scenario has been unambiguously verified only for B2~0648+27, whose \HI\ ring is gaseous tidal debris that is settling after a major merger occurred roughly 1.5 Gyr ago \citep[][see also Appendix \ref{app:hiproperties}]{emo06,emo08_0648}. For the other large-scale \HI\ discs in our sample, such a formation history is not immediately obvious; their host galaxies appear to have a more regular optical morphology without evidence for prominent stellar tidal features (Fig. \ref{fig:HIsample}) and they do not show evidence for a young stellar population as prominent as in B2~0648+27 across the bulge region \citep{emo06thesis}. Nevertheless, the surface brightness of the large-scale \HI\ discs/rings is probably too low for vigorous star formation to occur and hence they are likely to survive for many Giga-years. It is possible that the large, regular disc of B2 0258+35 and elliptical morphology of NGC~3894 and B2~1217+29 reflect more evolved stages in the evolution of a merger-system compared with B2~0648+27.\\
\vspace{-2mm}\\
{\sl Galaxy interactions:}\\
None of the radio galaxies in our sample show clear signs of ongoing gas-rich interactions with nearby companions. The only exception is the rare disc-dominated radio galaxy B2~0722+30, which shows that -- perhaps under specific circumstances -- a classical radio source can occur in a system that is undergoing gas-rich interactions \citep[][see also Sect. \ref{sec:seyferts}]{emo09}. Nevertheless, our \HI\ results indicate that low-power radio galaxies in general are not associated with violent, ongoing galaxy interactions that involve more than a few $\times 10^{8} M_{\odot}$ of \HI\ gas. In Sect. \ref{sec:environment} we already mentioned that smaller amounts of patchy \HI\ emission (such as the clouds of \HI\ emission observed in B2~1322+36) or slight distortions in the large-scale \HI\ discs could possibly be more subtle indications for less violent, gas-poorer or older galaxy interactions.\\
\vspace{-2mm}\\
{\sl Accretion of small companions:}\\
The presence of small amounts of \HI\ gas ($\la \rm few \times 10^{8} M_{\odot}$), for example in the case of B2~0055+30, could potentially be the result of the (continuous) accretion of gas from small companions. Such events are not likely to leave obvious observational evidence of the actual accretion event (or even in the total amount of \HI\ gas) at the sensitivity of our observations. From a much more sensitive study of \HI\ in early-type systems, \citet[][see Sect \ref{sec:radioquiet}]{oos09} suggest that accretion of cold gas -- likely over long time-scales -- may be a common feature among field early-type galaxies. They estimate, however, that the typical observable {\sl total} \HI\ accretion rate is smaller than 0.1 $M_{\odot}$ yr$^{-1}$ \citep[compared to at least 0.2 $M_{\odot}$ yr$^{-1}$ for field spiral galaxies;][]{san08}. We therefore argue that accretion of \HI\ gas from small companion galaxies does not provide a sufficient explanation for the formation of the large-scale \HI\ discs (with an \HI\ mass of several $\times 10^9 - 10^{10} M_{\odot}$) that we detected in our sample, because in that case either the number of events must be unphysically large, or the companion systems are large enough that the encounter would have resulted in a more violent galaxy-galaxy interaction or merger. It could, however, explain the presence of small clouds of \HI\ gas within the host galaxy, as for example detected in B2~0055+30. As mentioned in Sect. \ref{sec:environment}, estimates of the rates at which accretion of small gas-rich companions occurs in low-power radio galaxies can potentially be investigated by studying in detail the environment of nearby low-power radio galaxies (and comparing this with their radio-quiet counterparts), but this is beyond the scope of the current paper.\\
\vspace{-2mm}\\
{\sl Cold accretion of the IGM:}\\ 
\citet{ker05} show that gas from the inter-galactic medium (IGM) can be cooled along filamentary structures without being shock-heated, resulting in the accretion of cold gas onto the host galaxy. According to \citet{ser06}, the process of building a gaseous disc of about $10^{10} M_{\odot}$ through the process of cold accretion is certainly viable and takes many Gyrs. On smaller scales, \citet{kau06} show that through the cooling of hot halo gas, cold gas can be assembled onto a galactic disc. It thus seems possible that this cold accretion scenario is a potential process for forming -- over long timescales -- the range of \HI\ structures that we observe in our sample.\\
\vspace{0mm}\\
Of course, the galaxies in our sample are evolving continuously and it is certainly possible that a combination of the above mentioned mechanisms has occurred during their formation history. For example, the regular appearance of the \HI\ disc in B2~1217+29, combined with the typical elliptical morphology of the host galaxy, suggest that the system is old and that the \HI\ disc was created a long time ago. However, the two tails of \HI\ gas that stretch from either side of the disc \citep{mor06b} also suggest that the system is currently still accreting gas.

The possibility that large-scale \HI\ structures can be gradually assembled during the evolutionary history of early-type galaxies is supported through recent numerical simulations of `morphological quenching' by \citet{mar09}. They suggest that transformation from stellar discs to spheroids will stabilise the gas disc, quench star formation and create a red and dead early-type system while gas accretion continues. We argue that a stabilising factor (whether from the transformation from discs to spheroids or from the bulges of the galaxies themselves), but certainly also the low column densities of the gas, mean that the presence and ongoing accretion of large reservoirs of cold gas may occur naturally in early-type galaxies.

Despite the fact that the \HI\ emission properties can be used to investigate the formation history of the gas-rich host galaxies in our sample, it is good to keep in mind that for the majority of our sample sources (71$\%$) {\sl no} \HI\ emission-line structures have been detected. As we will see in the next Section, this \HI\ deficiency is particularly pronounced for the host galaxies of extended \FRI\ sources. In Sect. \ref{sec:nature} we will discuss in detail the nature of these \HI\ poor \FRI\ sources.

\subsection{The `H{\tiny \ }{\small I} mass - radio size' segregation}
\label{sec:relation}

As mentioned in Sect. \ref{sec:hiemission}, large amounts of \HI\ gas ($M_{\rm HI} \ga 10^9 M_{\odot}$) are only associated with the host galaxies of compact radio sources in our sample, while none of the host galaxies of the more extended \FRI\ radio sources shows similar amounts of large-scale \HI. A well known compact radio source from the literature that also contains a massive large-scale \HI\ disc (59 kpc in diameter and with $M_{\rm HI} = 1.5 \times 10^{10} M_{\odot}$ for $H_{0} = 71$ \kmsMp) is the nearby GPS source PKS~B1718-649 \citep[$P_{1.4 \rm GHz} = 24.2$ W Hz$^{-1}$;][]{ver95,tin97}. In \paperI, we already discussed the observed segregation in large-scale \HI\ mass content between compact and extended radio sources in our sample. It suggests that there is a physical link between the properties of the radio source and the presence of large-scale \HI\ structures. In this Section, we will review the possible explanations for the observed segregation, as discussed previously in \paperI.\\
\ \\
{\large {\sl Radio source ionisation/heating}}\\
\vspace{-2mm}\\
In \paperI\ we discarded the possibility that large-scale \HI\ discs/rings similar to those observed around our \HI-rich compact radio sources are fully ionised when the radio jets propagate outward. Although such a process may be viable when the radio jets are aligned in the plane of the disc \citep[as has been seen for Coma~A;][]{mor02}, such a chance alignment is not expected to occur frequently. Indications that radio jets propagating perpendicular to a large-scale \HI\ structure are not efficient in ionising the neutral gas on tens to hundreds of kpc scale come from the recent discovery of an enormous \HI\ disc in the nearby powerful radio galaxy NGC~612 \citep[][see also Sect. \ref{sec:NGC612}]{emo09} and from the presence of a 15 kpc wide central \HI\ disc in the vicinity of fast propagating radio continuum jets in Centaurus~A \citep[][see also Sect. \ref{sec:nature}]{cro09}. In addition, extensive emission-line studies show that \FRI\ radio galaxies generally do not contain features of ionised gas as extended, massive and regularly rotating as the \HI\ discs/rings that we find around a significant fraction of our compact radio sources \citep{bau88,bau89,bau92}. 

The situation may be different if the large-scale \HI\ discs/rings originated from the hot IGM that cooled and condensed onto the host galaxy in its neutral state (see Sect. \ref{sec:HIrich}). For X-ray luminous clusters and galaxy groups, \citet{bir04} and \citet{mcn05} showed that expanding X-ray cavities, produced by powerful radio jets that interact with the hot IGM, can in many cases quench cooling of this hot gas. \citet{bes06} argued from empirical evidence that in particular the moderately powerful radio sources (similar in power to the \FRI\ sources in our sample) are most effective in self-regulating the balance between cooling and heating of the hot gas surrounding these systems through recurrent activity. If this effect is strong enough and occurs also outside cluster environments, recurrent activity in extended \FRI\ radio galaxies may perhaps prohibit \HI\ structures from forming in the first place. An argument against this scenario could be the recent discovery of a very extended relic structure of radio continuum in our compact sample source B2~0258+35 (to be published in a forthcoming paper by Struve et al. [in prep.]). This indicates that the (recurrent) radio source in B2~0258+35 has not remained compact over the long time-scales required to build-up a large-scale \HI\ discs through cold accretion.\\
\smallskip\\
{\large {\sl Confinement}}\\
\vspace{-2mm}\\
A possible explanation for the observed segregation between \HI\ mass and radio source size is that the \HI-rich compact radio sources do not grow into extended sources because they are confined or frustrated by ISM in the central region of the galaxy. If the large amounts of \HI\ gas at large radii reflect the presence of significant amounts of gas in the central region \citep[e.g. as a result of a major merger, which both expels gas at large-scales and transports gas into the central kpc-scale region;][]{bar02}, the central ISM may be responsible for frustrating the radio jets if those are not too powerful. Interaction with the ambient medium has been suggested for each of the four most compact and most \HI-rich radio sources in our sample \citep{gir05a,tay98,gir05b}. Although it is not clear how much gas is needed to confine a radio source \citep*[see e.g. the discussion by][]{hol03}, \citet{gir05a,gir05b} argue that the relatively low power radio sources in NGC 4278 and B2 0648+27 cannot bore through the local ISM.\\
\smallskip\\
{\large {\sl Fuelling efficiency}}\\
\vspace{-2mm}\\
Alternatively, while large amounts of cold gas may provide sufficient material for fuelling the AGN, its distribution may be clumpy and the fuelling process may be inefficient. For example, while in a galaxy merger the geometry and the conditions of the encounter may be favourable to forming the observed large-scale \HI\ structures and even deposit significant amounts of gas in the central kpc-scale region, they may perhaps not be efficient in continuously channelling gas to the the very inner pc-scale region. This may prevent stable continuous fuelling of the AGN, so that large-scale radio structures do not develop. 

\citet*{sax01} argue that galaxy mergers can also temporarily interrupt the AGN fuelling process. They show that this likely happened in the nearest \FRI\ radio galaxy Centaurus~A (see Sect. \ref{sec:nature}), where the minor merger that formed the \HI\ disc also likely shut down the radio-AGN for a period of $\sim 10^{2}$ Myr, until re-started activity formed the compact inner radio lobes that we see today. It is also possible that the radio jets drive out substantial amounts of \HI\ gas from the centre [as observed in the nearby Seyfert galaxy IC~5063 \citep{oos00} as well as more powerful radio sources \citep{mor03apj593,mor05,mor05b}], terminating the fuelling process of the compact sources in our sample. 

\citet{gir05a} observed that the current radio source in B2~0258+35 displays variable levels of activity, suggestive of inefficient fuelling, and is therefore not expected to grow beyond the kpc-scale. As we will discuss in detail in Sect. \ref{sec:nature}, it has often been suggested that extended \FRI\ sources are fed through the accretion of hot circum-galactic gas. This likely results in a steady fuel supply, which allows these sources to grow to their large size before feedback effects may kick in \citep[e.g.][]{all06}.\\
\smallskip\\
\noindent In conclusion, we therefore argue that the observed `\HI\ mass - radio size' segregation in our sample is most likely the result of either confinement/frustration of compact radio jets by a central dense ISM, or inefficient fuelling of a significant fraction of compact jets compared with a more steady fuelling of extended \FRI\ sources. It is very well conceivable that a combination of both processes is at work in an environment where re-started radio sources continuously have to try to fight their way through a dense ISM until the fuelling process is temporarily halted.\\
\smallskip\\
\indent A similar segregation in \HI\ mass with radio source size is found for high-$z$ radio galaxies by \citet{oji97}. While they detect strong \HI\ absorption in the majority of the galaxies' Ly$\alpha$ halos, they also find that 90$\%$ of the smaller ($<$50 kpc) radio sources have strong associated \HI\ absorption, whereas only a minority of the more extended sources contain detectable \HI\ absorption. Van Ojik et al. prefer the explanation that these small radio sources reside in dense, possibly (proto) cluster environments, where large amounts of neutral gas can exist and where the radio source vigorously interacts with the ambient gaseous medium (although also other possible scenarios are discussed). Although the radio galaxies in our sample are much less powerful and were selected {\sl not} to lie in dense cluster environments, it is nevertheless intriguing that we find a similar segregation in \HI\ content between compact and extended sources in the nearby Universe.

\subsection{Comparison radio-quiet early-type galaxies}
\label{sec:radioquiet}

Over the past two decades, case studies of early-type galaxies have imaged large-scale \HI\ structures associated with these systems \citep[see e.g][]{dri88,dri89,sch94,sch95,mor97,gor97,sad00,oos01,oos02,ser06,don09}. Many of these large-scale \HI\ structures have an \HI\ mass and morphology similar to the structures that we find around the \HI-rich radio galaxies in our sample. Even a significant fraction of early-type galaxies that optically can be classified as `dry' merger systems (i.e. systems that supposedly formed as a results of a major merger between red and dead galaxies without any gaseous component), is found to contain significant amounts of cool gas when observed with radio telescopes \citep{don07,ser10}.

Recently, \citet{oos07,oos09} have completed two studies that obtained quantitative results on the occurrence and morphology of large-scale \HI\ in early-type galaxies (not selected on radio loudness). These two studies are therefore ideally suited for a detailed comparison with the \HI\ properties of our complete sample of B2 radio galaxies. The first study by \citet{oos07} involved follow-up imaging of \HI\ in early-type galaxies detected by \citet{sad01} in the single-dish \HI\ Parkes All-Sky Survey \citep[HIPASS;][]{bar01HIPASS,mey04}. The second study by \citet{oos09} involved deep \HI\ imaging of 33 nearby early-type galaxies selected from a representative sample of early-type galaxies observed with the optical integral field spectrograph SAURON. Of these 33, 20 are field early-type galaxies \citep[an extension of earlier work done by][]{mor06b}, while 13 are Virgo-cluster systems. Since we excluded cluster sources from our B2 radio galaxy sample (Sect. \ref{sec:sample}), we will only take into account the field early-type galaxies from the SAURON sample in the remainder of this Section. The early-type galaxies from the HIPASS and SAURON samples have a typical radio power $P_{\rm 1.4 GHz} < 10^{22}$ W, i.e. significantly lower than that of our B2 sample of radio galaxies. The B2 and the SAURON sample have one source in common, namely B2~1217+29, which is by far the strongest radio source in the SAURON sample and the second weakest source in our B2 sample. It is interesting to note that the only two objects in the HIPASS sample with $P_{\rm 1.4 GHz} > 10^{22.1}$ W both have a {\sl compact} radio source as well as significant amounts of large-scale \HI\ gas, in agreement with the trend that we find in our B2 sample (Sect. \ref{sec:relation}).

\begin{table}
\centering
\caption{\HI\ detection rates of the various samples of early-type galaxies}
\label{tab:detectionrates}
\begin{tabular}{l|c|c|c}
 & HIPASS & B2 & SAURON \\
 &        &    & (field sample) \\
\hline
\hline
$\#$ galaxies & 818 & 21$^{*}$ & 20 \\
detection limit ($M_{\odot}$) & $\sim 10^{9}$ & ${\rm few} \times 10^{8}$ & ${\rm few} \times 10^{6}$ \\
detection rate ($\%$) & 9$^{\dag}$ & 29 & 70 \\
\hline
$\%$ with $M_{\rm HI} > 10^{9} M_{\odot}$ & 9$^{\dag}$ & 10 & 10 \\
$\%$ with $M_{\rm HI} \ga 10^{8} M_{\odot}$ & - & 29 & 35 \\
\hline
\hline
\end{tabular}\\
\vspace{2mm} 
\flushleft{{\small
$^{*}$ Complete B2 sample, does not include NGC~3894 and B2~1557+26 (see Sect. \ref{sec:sample}).\\
$^{\dag}$ From Sadler et al. (in prep.); for early results see \citet{sad01}.}}
\end{table}

Table \ref{tab:detectionrates} summarises both the \HI\ detection limits and the \HI\ detection rates of the HIPASS, B2 and SAURON samples. There is a substantial difference in sensitivity between the three samples, which makes a comparison of the \HI\ detection rates difficult. Nevertheless, when looking at the high-mass end, the percentage of sample sources with $M_{\rm HI} \ga 10^{9} M_{\odot}$ is roughly the same for the three samples. Towards the low-mass end, the percentage of radio-quiet galaxies in the SAURON field sample with $M_{\rm HI} \ga 10^{8} M_{\odot}$ is also very similar to that of our B2 sample of radio-loud galaxies. In the presence of a radio continuum source, low amounts of \HI\ gas could be observed in absorption rather than emission. When including the additional three galaxies from our B2 sample with a tentative \HI\ absorption detection, the \HI\ detection rate for our sample of radio-loud early-type galaxies raises to 43$\%$. Therefore -- within the significant uncertainty due to non-uniform sensitivity and relatively low number statistics -- there does not appear to be a significant difference in the \HI\ total mass content between the radio-loud and radio-quiet samples.\footnote{We note again that these detection rates are based on non-cluster early-type galaxies; \citet{dis07} and \citet{oos09} show that the \HI\ detection rate of early-type galaxies in the Virgo Cluster is dramatically lower.}

Regarding the morphology, about two-thirds of the \HI\ structures that are imaged in the HIPASS follow-up study are large and regularly rotating discs or rings \citep{oos07}, similar to the \HI\ structures at the high-mass end of the B2 sample. The morphology of the \HI\ structures in the SAURON sample is diverse, with \HI\ morphologies ranging from regular rotating discs to irregular clouds, tails and complex distributions. Also here, the strongest \HI\ detections are often regular disk/ring-like structures, although the good sensitivity of these observations clearly reveals more complex kinematics than observed for the HIPASS and B2 samples \citep[][]{mor06b,oos09}. At the low-mass end, the \HI\ structures in the SAURON sample often have a much more irregular or clumpy appearance (as do B2~1322+36 and B2~0055+30 in our radio-loud B2 sample).

Thus -- as far as we can tell from the limited comparison between the three systematic studies -- there appears to be no major difference in both \HI\ detection rate and \HI\ morphology between the radio-quiet and radio-loud early-type galaxies in these samples. For sure, across the range of masses that we studied in this paper, there is no evidence that our radio-loud sample has a higher content of large-scale \HI\ gas or contains more tidally distorted \HI\ structures than the radio-quieter samples. {\sl If confirmed by larger samples with comparable sensitivity, this may indicate that the radio-loud phase could be just a short period that occurs at some point during the lifetime of many -- or maybe even all? -- early-type galaxies.} This would add to the growing evidence that radio-AGN activity can be an episodic or recurrent phenomenon \citep[see e.g.][for a review]{sai10}.

These conclusions are in agreement with a recent study of CO in nearby radio galaxies by \citet{oca10}, who find no difference in the molecular hydrogen (H$_{2}$) mass content between their sample of nearby radio galaxies and a sample of genuine early-type galaxies by \citet*{wik95}. Our results also agree with the fact that \citet{bet01} and \citet{bet09} find that -- regarding low-power radio AGN -- both radio and non-radio ellipticals follow the same Fundamental Plane and Core Fundamental Plane. \citet{cap06} \citep[also][]{cap05,bal06} show that radio-loud AGN occur only in early-type galaxies with a shallow inner cusp (`core-galaxies'), while those with steep (power-law) cusps solely harbour AGN that are radio-quiet. In that case, a radio-loud phase could be a common feature only among `core' early-type galaxies. Nevertheless, \citet{cap06} also show that -- apart from the properties of the central cusp -- this radio-loud/radio-quiet dichotomy is not apparently related to other properties of the host galaxy.

\subsection{The nature of low-power radio galaxies}
\label{sec:nature}

\subsubsection{\FRI\ sources}

The lack of detectable amounts of \HI\ in most \FRI\ radio galaxies is in agreement with the growing evidence that the AGN in these systems are {\sl not} associated with galaxy mergers, collisions or violent ongoing interactions that involve significant amounts of cool gas. While this was already suggested from optical studies by \citet{hec86} and \citet{bau92} (see Sect. \ref{sec:intro}), our \HI\ results provide -- for the first time in a systematic way -- {\sl direct} evidence for the lack of cold gas that would be associated with such violent events. Various studies suggest that low-power \FRI\ radio sources generally also lack evidence for a thick torus and classical accretion disc \citep{chi99,mor01}. Furthermore, there is growing evidence that these low-power radio sources are likely fed through a quasi-spherical accretion of hot gas from the galaxy's halo or IGM directly onto the nucleus \citep{bes06,all06,balma08}. As we already mentioned in Sect. \ref{sec:intro}, \citet{har07} agree with such a scenario, but extend on this idea in the sense that all low-excitation AGN (including almost all - but not exclusively - \FRI\ sources) may share this accretion mechanism \citep[see also][]{bal08}. We argue that the lack of large amounts of \HI\ gas in extended \FRI\ sources, as well as the similarity in \HI\ properties between our sample of low-power radio galaxies and radio-quiet(er) early-type galaxies, is in agreement with the growing evidence that the AGN in \FRI\ radio galaxies is generally fed through the steady accretion of hot circum-galactic gas.

Although accretion of hot IGM directly onto the central engine provides a good explanation for the \HI\ properties of \FRI\ radio galaxies, other possible feeding mechanisms need to be considered. As mentioned in Sect. \ref{sec:HIrich}, our observations cannot rule out that much less violent, gas-poor or old interactions may be associated with \FRI\ galaxies. \citet{col95} argue that elliptical-elliptical mergers -- often referred to as `dry' mergers -- may occur frequently among \FRI\ sources. Since a mass accretion rate as little as $10^{-3} - 10^{-5} M_{\odot}$ yr$^{-1}$ may be sufficient to power a radio source \citep[e.g.][]{gor89}, even a relatively dry merger (which does not contain observable amounts of \HI\ gas) could potentially still carry enough fuel to feed the radio source for a significant time. Given these low mass accretion rates, it may even be conceivable that stellar mass-loss processes \citep[e.g.][]{wil00} are able to deliver the potential AGN-fuel to the central region.

It may also be possible that, over long time-scales, (continuous) accretion of gas can build up a concentration or even a disc of gas and dust in the central region \citep[which are known to exist in \FRI\ radio galaxies;][]{ver99,cap00,rui02}, which may potentially provide the fuel supply for the AGN. As mentioned in Sect. \ref{sec:HIrich}, either cold accretion from the IGM or minor accretion events of companion galaxies (which do not leave observable amounts of \HI\ debris) are potential processes that may drive this accretion. 

\citet{oos09} find an intriguing trend that `normal' early-type galaxies that are detected in \HI\ (but often with an \HI\ mass lower than the detection limit in our sample) are more likely to contain a very faint and (in many cases) very compact radio continuum component compared with early-type galaxies that are not detected in \HI. This suggests that the cold gas contributes -- at least to some extent -- to the feeding of a very low-power radio-AGN in some early-type galaxies. We note, however, that the radio continuum sources in these systems are several orders of magnitude less powerful than the classical `low-power' radio sources in our B2 sample, hence it is very well conceivable that there are substantial differences between these two types of AGN (though a more detailed comparison certainly deserves further attention).

We therefore conclude that our \HI\ results are in agreement with the growing evidence that classical \FRI\ radio sources are fed through the steady accretion of hot circum-galactic gas and {\sl not} by violent gas-rich galaxy mergers and interactions, but that there are other possible mechanisms that cannot be ruled out.\\
\vspace{2mm}\\
{\it Centaurus~A}\\
\vspace{-2mm}\\
The lack of detectable amounts ($\ga \rm few \times 10^{8} M_{\odot}$) of large-scale \HI\ in nearby \FRI\ radio galaxies seems, at first sight, in contradiction with \HI\ observations of Centaurus~A. Cen~A is by far the nearest \FRI\ radio galaxy and hence studied in much greater detail than any other radio galaxy. Cen~A has an extended (650 kpc) \FRI\ radio source and contains a total of $6 \times 10^8 M_{\odot}$ of \HI\ gas \citep[$4.5 \times 10^8 M_{\odot}$ in a central 15 kpc disc and $1.5 \times 10^8 M_{\odot}$ in faint outer shells;][]{gor90,sch94,str09}. From X-ray observations, \citet{kra03} and \citet{cro07,cro09} argue that Cen~A may be a non-typical low-power radio galaxy in that it shares some properties (supersonic expansion of one of the inner radio lobes and a high intrinsic absorption of the nucleus) with more powerful \FRII\ radio galaxies, which are often associated with gas-rich galaxy mergers. Indeed, it has been argued that the \HI\ structures in Cen~A formed as a result of a minor merger \citep{sch94}, so it is certainly possible that Cen~A is indeed not a `typical' \FRI\ radio galaxy. 

On the other hand, \citet[][]{str09} find no evidence from the \HI\ properties that the minor merger event in Cen~A is responsible for fuelling the current episode of radio-AGN activity \citep[in fact,][\ suggest that this minor merger was responsible for temporarily shutting down the radio-AGN rather than triggering the current episode of radio-AGN activity]{sax01}. \citet{str09} also show that, if Cen~A would be located at the average distance of our B2 sample sources, a significant part of the \HI\ disc would not be detectable in emission but in absorption. In addition, at large distances the relatively compact continuum from the inner radio lobes is likely to dominate the radio-source structure. 

From the \HI\ results presented in this paper, there is thus no unambiguous evidence either in favour or against the idea that Cen~A is not a typical \FRI\ radio galaxy, but it does serve as a strong reminder of the observational limitations of our current sample.

\subsubsection{Low-power compact sources}

Contrary to the extended \FRI\ sources, a significant fraction of the low-power compact sources in our sample do contain enormous discs/rings of \HI\ gas, some of which could be related to a past gas-rich merger event. However, we saw in Sect. \ref{sec:HIrich} that these \HI\ structures are at least one to several Gyr old. The lifetime of extended, low-power radio sources is generally believed to be not more than about $10^8$ yr \citep[e.g.][]{par02} and the compact radio sources in our sample are believed to be even significantly younger \citep{gir05a,tay98}. This suggests that the onset of the current episode of radio-AGN activity started long after the initial formation of these \HI\ discs. In \citet{emo06} (where we studied the case of B2~0648+27) we discussed that, in (post-)merger systems, significant time-delays between the initial merger and the onset of the radio-AGN activity may not be uncommon. Also, it is possible that there have been previous episodes of AGN activity -- as we mentioned in Sect. \ref{sec:radioquiet}, growing evidence suggests that AGN activity could be episodic in nature \citep[e.g.][]{sai10} and there are indications that there is likely a high incidence of \HI\ absorption associated with rejuvenated radio sources \citep{sai07,cha10}. Nevertheless, a direct causal connection between the formation of the \HI\ discs/rings and the triggering of the {\sl current} episode of radio-AGN activity is not immediately apparent.

Therefore, the feeding mechanism of these compact radio sources remains ambiguous. However, the `\HI\ mass - radio size' segregation that we find (Sect. \ref{sec:HIrich}) indicates that the fuelling mechanism and/or the evolution of these \HI-rich compact radio sources is fundamentally different from that of the extended \FRI\ sources and somehow related to the presence of large amounts of \HI\ gas.

\subsubsection{Comparison with Seyfert sources}
\label{sec:seyferts}

Our \HI\ results on low-power radio galaxies are also interesting when compared with the properties of nearby Seyfert galaxies. Seyfert nuclei are often found in spiral galaxies and a significant fraction contains a very compact and low luminosity radio-AGN \citep[with a total radio power well below that of the low-power radio galaxies in our sample, e.g.][]{ho01}. Thus, while disc-dominated Seyfert galaxies (with a prominent \HI\ and stellar disc) often contain a very faint radio-AGN, we find that a significant fraction of much brighter compact radio sources is hosted by early-type galaxies with more diffuse large-scale \HI\ discs and that classical, extended \FRI\ sources occur in early-type galaxies without a prominent disc. This may hint to a continuum in radio-source properties from late- to early-type galaxies. A more detailed study on the comparison between radio-AGN activity and the host galaxy's disc properties across the full spectrum of galaxy morphologies deserved further investigation, but is beyond the scope of this paper.

\citet{kuo08} and \citet{tan08} found from \HI\ studies that local disc-dominated Seyfert galaxies generally show evidence for ongoing gas-rich interactions and that these interactions are important for the occurrence of the nuclear activity. From the lack of evidence of ongoing gas-rich interactions among our sample sources, we argue that the fuelling mechanism of low-power radio galaxies is likely fundamentally different from that of disc-dominated Seyfert galaxies. We note, however, that the Seyfert sources in the samples of \citet{kuo08} and \citet{tan08} all have a redshift comparable to the low-redshift range of our sample of radio galaxies, hence a more in-depth investigation would be necessary in order to determine to what extent sensitivity issues limit this comparison.

Interestingly, the only disc-dominated \FRI\ radio galaxy in our sample (B2~0722+30) shows \HI\ properties similar to those of local disc-dominated Seyfert systems (namely a regular \HI/stellar disc and \HI-rich interactions with companions). This could mean that the host galaxy environment and AGN feeding mechanism of B2~0722+30 more closely resembles that of nearby Seyfert galaxies rather than that of low-power radio galaxies in general \citep[despite the clear evidence for a typical \FRI\ radio-AGN not commonly observed among Seyferts; see][]{emo09}.

\subsubsection{Comparison with powerful (\FRII) sources}
\label{sec:NGC612}

As mentioned in Sect. \ref{sec:intro}, powerful radio galaxies with strong emission-lines have -- in contrast to our results on \FRI\ sources -- often been associated with gas-rich galaxy mergers or collisions. We recently found evidence for a large-scale (140 kpc) \HI\ disc associated with the nearby powerful radio galaxy NGC~612 \citep[PKS~0131-36;][]{emo08_NGC612}. This radio source has clear \FRII\ properties and shows a faint \HI\ bridge that stretches across 400 kpc toward a gas-rich companion galaxy, indicating that a collision between both systems likely occurred \citep{emo08_NGC612}. In a future paper we will investigate the large-scale \HI\ properties of a small sample of nearby powerful \FRII\ radio galaxies, which will allow us to compare the general \HI\ properties between low- and high-power (as well as low- and high-excitation) radio galaxies.

\section{Conclusions}
\label{sec:conclusions}

From our study of large-scale \HI\ in a complete sample of nearby low-power radio galaxies (compact and \FRI), we derive the following conclusions:\\
\vspace{-1mm}\\
{\sl i).} Our detection rate of \HI\ emission directly associated with the radio galaxy is 29$\%$ (with a detection limit of $\sim 10^8 M_{\odot}$);\\
\vspace{-3mm}\\
{\sl ii)} We find {\sl no} evidence for {\sl ongoing} gas-rich galaxy mergers, collisions or violent interactions associated with the early-type host galaxies of low-power radio sources. At the high-mass end, all the \HI\ structures are fairly regularly rotating large-scale discs/rings, while at the low-mass end (several $\times 10^{7} M_{\odot}$) the \HI\ distribution appears much more clumpy. The large-scale \HI\ discs/rings are at least one to several Gyr old;\\
\vspace{-3mm}\\
{\sl iii).} There is a clear segregation in \HI\ mass content between compact and extended radio sources in our sample. Large amounts of \HI\ (with $M_{\rm HI} \ga 10^{9} M_{\odot}$) are only observed around host galaxies with a compact radio source, while none of the host galaxies of the more extended \FRI\ radio sources shows similar amounts of large-scale \HI. This suggests that there is a physical link between the properties of the radio source and the presence of large-scale \HI\ structures, which we ascribe most likely to either confinement/frustration of the compact radio sources by the presence of large amounts of gas, or to the lack-of-growth of the compact sources as a result of inefficient fuelling;\\
\vspace{-3mm}\\
{\sl iv).} Our \HI\ results indicate that extended \FRI\ radio galaxies are generally hosted by \HI-poor galaxies. Only low amounts of \HI\ ($< 10^8 M_{\odot}$) have been detected in a small fraction of these systems. These results are in agreement with the growing belief that extended \FRI\ radio galaxies are fuelled through the accretion of their circum-galactic hot gas (although other mechanisms cannot be excluded);\\
\vspace{-3mm}\\
{\sl v).} From a limited comparison with samples of radio-quiet early-type galaxies, our complete sample of low-power radio galaxies shows no apparent difference in \HI\ properties (detection rate, mass and morphology) compared with these radio-quiet samples. If confirmed by larger samples with uniform sensitivity, this could mean that a classical low-power radio source may occur at some point during the lifetime of many -- or perhaps even all -- early-type galaxies (at least the ones with a shallow central cusp).\\

\section*{Acknowledgments}

We would like to thank Jacqueline van Gorkom for her great help and useful discussions. Also many thanks to our referee Dhruba Saikia for valuable suggestions that improved this paper. BE thanks Columbia University, the Kapteyn Astronomical Institute and ASTRON for their hospitality during parts of this project and acknowledges the corresponding funding received from the University of Groningen and the Netherlands Organisation for Scientific Research - NWO (Rubicon grant 680.50.0508). The National Radio Astronomy Observatory is a facility of the National Science Foundation operated under cooperative agreement by Associated Universities, Inc. The Westerbork Synthesis Radio Telescope is operated by the ASTRON (Netherlands Foundation for Research in Astronomy) with support from NWO. The Michigan-Dartmouth-MIT Observatory at Kitt Peak is owned and operated by a consortium of the University of Michigan, Dartmouth College, Ohio State University, Columbia University and Ohio University. The NASA/IPAC Extragalactic Database (NED) is operated by the Jet Propulsion Laboratory, California Institute of Technology, under contract with the National Aeronautics and Space Administration.

\bibliographystyle{mn} 
\bibliography{auth_total_HInew4} 

\appendix

\section{Individual \HI\ sources}
\label{app:hiproperties}

This Appendix gives a detailed description of the individual sources in our sample for which \HI\ has been detected in emission and/or absorption (see Sect. \ref{sec:results}).\\
\vspace{-2mm}\\
{\bf B2 0055+30 (NGC~315):} \HI\ results of this sources have been published by \citet{mor09_NGC315}. The absorption profile contains a broad component slightly redshifted from the systemic velocity as well as a narrow component redshifted by about 460 \kms\ \citep[see also][]{hec83}. \citet{mor09_NGC315} favour the idea that the broad component may represent gas that is falling into the nucleus, while the narrow component is likely an \HI\ cloud at larger distance from the centre. A small cloud of \HI\ emission is also detected within the host galaxy at roughly the same velocity as the narrow absorption component \citep{mor09_NGC315}. B2~0055+30 is the most extended \FRI\ radio source in our sample and has an asymmetric jet/lobe structure with a peculiar bend at one end \citep[see][and references therein]{lai06}.\\
\vspace{-2mm}\\
{\bf B2 0222+36:} The detection of \HI\ absorption in this galaxy with a fairly compact radio source is tentative and needs to be confirmed with additional observations.\\
\vspace{-2mm}\\
{\bf B2 0258+35:} The \HI\ results of this source will be published in detail in a forthcoming paper by Struve et al. (in prep.). The \HI\ emission-line gas around B2 0258+35 is distributed in a regularly rotating disc with a diameter of 160 kpc \citep*[see][for a position-velocity plot]{str08_0258}. A slight asymmetry appears in the \HI\ gas towards the outer western part of the otherwise settled disc. It is likely that the bulk of the absorbing \HI\ gas is located in the large-scale disc, although part of it could also come from a circum-nuclear disc \citep[see][]{str08_0258}. Besides several \HI\ companions outside the \HI\ disc, our deep optical image shows what appears to be a very faint and tidally disrupted system at the northern edge of the \HI\ disc, most prominently visible around RA=03h01m48s, dec=+35$^{\circ}$15$^{\prime}$15$^{\prime\prime}$ (this feature will be described in more detail by Struve et al [in prep.]). The optical host galaxy has a large bulge component and what appears to be a very faint and tightly wound spiral disc. Our \HI\ and optical data are in good agreement with earlier data presented by \citet{noo05}. The radio source in B2~0258+35 has been classified as Compact Steep Spectrum \citep[CSS; ][]{san95}.\\
\vspace{-2mm}\\
{\bf B2 0648+27:} We studied this galaxy in great detail in \citet{emo06,emo08_0648}. The \HI\ gas is distributed in a massive, regularly rotating ring-like structure with a diameter of 190 kpc. Deep optical imaging shows a distorted optical morphology and a faint stellar tail or partial ring that follows the \HI\ ring \citep{emo08_0648}. The stellar light across the host galaxy is dominated by a 0.3 Gyr post-starburst stellar population \citep{emo06}. We argued that B2~0648+27 formed from a major merger event roughly 1.5 Gyr ago, after which \HI\ gas that was expelled from the system during the merger had time to fall back and settle around the host galaxy \citep{emo06}. The radio source is compact, with a minimum estimated age of only about 1 Myr \citep{gir05a}. The current phase of radio-AGN activity has therefore started late in the lifetime of the merger.\\
\vspace{-2mm}\\
{\bf B2 0722+30:} We studied this galaxy in great detail in \citet{emo09}. B2~0722+30 is the only late-type galaxy in our sample. The regularly rotating \HI\ emission follows the edge-on stellar disc. Part of the \HI\ disc is seen in absorption and not taken into account in the total intensity image of Fig. \ref{fig:HIsample} and \HI\ mass estimate in Table \ref{tab:hiradiogalaxies}. B2~0722+30 has an \HI-rich environment, with gas-rich galaxies that are in ongoing interaction \citep[see][]{emo09}. The radio source reaches beyond the optical boundary of the host galaxy and has an \FRI\ morphology \citep{fan86}. It is extremely rare for disc galaxies to host a classical radio source \citep[see][]{ver01,kee06,emo08_NGC612,emo09}, hence we warn the reader that B2~0722+30 should be regarded as a special case in our sample.\\
\vspace{-2mm}\\ 
{\bf B2 1217+29 (NGC~4278):} \HI\ observations of this sources have been published by \citet{rai81}, \citet{lee94} and \citet{mor06b}. They all detect an \HI\ disc with regular rotation, although the gas is not co-planar with the rotation of the stars in the inner part of the galaxy and shows indications for non-circular motions. Deep \HI\ imaging by \citet{mor06b} shows that the \HI\ disc is somewhat asymmetric and slightly elongated eastward (roughly in the direction of a close companion) and contains two faint tails of \HI\ gas on either side. Interestingly, B2~1217+29 shows no evidence for \HI\ absorption against the central continuum. In our deep optical image, both B2~1217+29 and the smaller, close companion towards the north-east have the appearance of a typical elliptical galaxy, although B2~1217+29 contains a faint dust-lane stretching from north-east to west of the nucleus. Typical for early-type galaxies, B2~1217+29 contains a relatively old stellar population \citep[e.g.][]{san06}. The radio source B2 1217+29 is compact and the second weakest source in our sample (see Table \ref{tab:sourceproperties}).\\
\vspace{-2mm}\\
{\bf B2 1321+31:} Two tentative (3$\sigma$) absorption features are detected against the radio continuum of B2~1321+31; one against the central radio continuum \citep*[slightly redshifted with respect to the optical systemic velocity as determined by][]{woo06} and another against the bright radio continuum at the tip of the north-western lobe, roughly 100 kpc from the nucleus (see Fig \ref{fig:absorption1321_extended}). The tentative \HI\ feature against this outer lobe is spatially unresolved, since the estimated \HI\ column density of the absorbing gas ($N_{\rm HI} \sim 3.6 \times 10^{21}$ cm$^{-2}$) would have been sufficiently high for detecting part of this \HI\ structure also in emission outside the radio continuum, which is not the case. There is no galaxy visible near the location of this outer absorption in optical SDSS images. The tentative absorption against the outer lobe of B2~1321+31 resembles \HI\ absorption features detected against the outer edge of the powerful radio sources 3C 234 \citep{pih01} and 3C 433 \citep{mor04} and could represent a region where the radio plasma interacts with ambient inter-galactic medium. The extended radio source has a typical \FRI\ morphology.\\ 
\vspace{-2mm}\\
{\bf B2 1322+36 (NGC~5141):} This system shows two clouds of \HI\ emission in the direction of the nearby companion galaxy NGC 5142. As can be seen from Fig. \ref{fig:HIsample}, \HI\ absorption detected against the radio continuum is also slightly extended in this direction. The column density of the absorbing \HI\ gas is very similar to the peak column density seen in emission. This suggests that the emission and the absorption are possibly part of the same large-scale \HI\ structure, which has a too low surface brightness to be detected in emission at other locations. The systemic velocity based on the stellar kinematics \citep[see][]{noe03} corresponds to the peak in the \HI\ absorption profile. We detect no optical counterpart at the location of the \HI\ emission in our deep optical image. B2~1322+36 shows no obvious features in our deep optical image, but the bulge dominated companion system NGC~5142 shows indications of a minor and very faint warped disc. The radio source has a total linear extent of 19 kpc and a typical \FRI\ morphology.\\ 
\vspace{-2mm}\\
{\bf B2 1447+27:} The detection of \HI\ absorption in this system is very marginal, and only seen in a data cube with robust or natural weighting. For a uniform weighted cube the absorption feature disappears in the noise. Additional observations are necessary to verify this tentative detection. It is not clear whether this absorption represents gas in the very nuclear region or at larger scales. The radio source in B2~1447+27 is compact.\\
\vspace{-2mm}\\
{\bf NGC 3894:} The \HI\ around NGC~3894 is distributed in what appears to be an edge-on ring-like (or possibly a disc-like) structure. Our deep optical image reveals a faint but extended dust lane along the direction of the \HI\ ring. The \HI\ ring seems distorted at the location of the nearby barred spiral galaxy NGC~3895, which lies 27 kpc east-north-east of NGC~3894 and does not contain any observable \HI. The \HI\ absorption against the unresolved radio source in NGC 3894 has a clear double-peaked profile \citep[see also][]{dic86,gor89,pec98,gup06}. We argue that the bulk of the absorbing gas is likely part of the large-scale \HI\ ring. As shown in \paperI\ and \citet{emo06thesis}, NGC~3894 is located in an environment of several nearby \HI-rich galaxies. The radio source is compact \citep[][]{tay98}.

\end{document}